    \documentclass[
        fleqn,
        usenatbib,
    ]{mnras}
    \usepackage{
        amsmath,
        amssymb,
        newtxtext,
        newtxmath,
        ae, aecompl,
        graphicx,
        booktabs,
        xcolor,
    }

    \newcommand*{\rd}[2]{\frac{\mathrm{d}#1}{\mathrm{d}#2}}

    \newcommand*{\abs}[1]{\left|#1\right|}
    \newcommand*{\ev}[1]{\left\langle#1\right\rangle}
    \newcommand*{\p}[1]{\left(#1\right)}
    \newcommand*{\s}[1]{\left[#1\right]}
    \newcommand*{\z}[1]{\left\{#1\right\}}
    \newcommand*{\bm}[1]{\mathbf{#1}}
    \newcommand*{\uv}[1]{\hat{\boldsymbol{\mathbf{#1}}}}

    \let\Re\undefined
    \let\Im\undefined
    \DeclareMathOperator{\Im}{Im}
    \DeclareMathOperator{\Re}{Re}

    \newlength{\colummwidth}
\title[Multi-planetary Spin Equilibria]{Dynamics of Colombo's Top:
Non-Trivial Oblique Spin Equilibria of Super-Earths in Multi-planetary Systems}
\author[Y. Su and D. Lai.]{
Yubo Su,$^1$\thanks{E-mail: yubosu@astro.cornell.edu},
Dong Lai$^1$
\\
$^1$ Cornell Center for Astrophysics and Planetary Science, Department of
Astronomy, Cornell University, Ithaca, NY 14853, USA\\
}

\date{Accepted XXXX\@. Received YYYY\@; in original form ZZZZ.}

\pubyear{2021}

\begin{document}\label{firstpage}
\pagerange{\pageref{firstpage}--\pageref{lastpage}}
\maketitle

\begin{abstract}
Many Sun-like stars are observed to host close-in super-Earths (SEs) as part of
a multi-planetary system. In such a system, the spin of the SE evolves due to
spin-orbit resonances and tidal dissipation. In the absence of tides, the
planet's obliquity can evolve chaotically to large values. However, for close-in
SEs, tidal dissipation is significant and suppresses the chaos, instead driving
the spin into various steady states. We find that the attracting steady states
of the SE's spin are more numerous than previously thought, due to the discovery
of a new class of ``mixed-mode'' high-obliquity equilibria. These new equilibria
arise due to subharmonic responses of the parametrically-driven planetary spin,
an unusual phenomenon that arises in nonlinear systems. Many SEs should
therefore have significant obliquities, with potentially large impacts on the
physical conditions of their surfaces and atmospheres.
\end{abstract}


\section{Introduction}\label{s:intro}

The obliquity of a planet, the angle between the spin and orbital axes, plays an
important role in the atmospheric dynamics, climate, and potential habitability
of the planet. For instance, the atmospheric circulation of a planet changes
dramatically as the obliquity increases beyond $54^\circ$, as the averaged
insolation at the poles becomes greater than that at the equator
\citep{ferreira2013, lobo2020atmospheric}. Variations in insolation over long
timescales can also affect the habitability of an exoplanet
\citep{spiegel2010generalized, armstrong2014effects}. In the Solar System,
planetary obliquities range from nearly zero for Mercury and $3.1^\circ$ for
Jupiter, to $23^\circ$ for Earth and $26.7^\circ$ for Saturn, to $98^\circ$ for
Uranus. The obliquities of exoplanets are challenging to measure, and so far
only loose constraints have been obtained for the obliquities of faraway
planetary-mass companions \citep{bryan2020obliquity, bryan2021obliquity}.
Nevertheless, there are prospects for better constraints on exoplanetary
obliquities in the coming years \citep{snellen2014fast, bryan2018constraints,
seager2002constraining}. Substantial obliquities are of great theoretical
interest for their proposed role in explaining peculiar thermal phase curves of
transiting planets \citep{millholland_signatures, ohno_infer_obl} and various
other open dynamical puzzles \citep{millholland_wasp12b,
millholland2019obliquity, millholland2020formation, su2021dynamics}.

The obliquity of a planet reflects its dynamical history. Some obliquities could
be generated in the earlest phase of planet formation, when/if proto-planetary
disks are turbulent and twisted \citep{tremaine1991, jennings2021turbulent}.
Large obliquities are commonly attributed to giant impacts or planet collisions
as a result of dynamical instabilities of planetary orbits \citep{original_gi,
benz1989tilting, korycansky1990one, dones1993does, morbidelli_gi,
li2020planetary, li2021giant}. Repeated planet-planet scatterings could also
lead to modest obliquities \citep{hong2021, gongjieli2021}. Substantial
obliquity excitation can be achieved over long (secular) timescales via
spin-orbit resonances, when the spin precession and orbital (nodal) precession
frequencies of the planet become comparable \citep{ward2004I, ward2004II,
ward_jupiter, vokrouhlicky2015tilting, millholland_disk, saillenfest2020future,
su2020, saillenfest2021large}. Such resonances are likely responsible for the
obliquities of the Solar System gas giants and may have also played a role in
generating obliquities of ice giants \citep{hamilton_tilting_ice}.  For
terrestrial planets, multiple spin-orbit resonances and their overlaps can make
the obliquity vary chaotically over a wide range \citep{laskar1993chaotic,
touma1993chaotic, correia2003long}.

A large fraction ($30-90\%$) of Sun-like stars host close-in super-Earths (SEs),
with radii $1-4R_{\oplus}$ and orbital distances $\lesssim 0.5 \;\mathrm{AU}$,
mostly in multi-planetary ($\geq 3$) systems
\citep[e.g.][]{lissauer2011architecture, howard2012planet, zhu201830,
sandford2019multiplicity, yang2020, he2021}. In these systems, the SE orbits
are mildly misaligned with mutual inclinations $\sim 2^\circ$
\citep{lissauer2011architecture, tremaine2012, fabrycky_mutual_incs}, which
increase as the number of planets in the system decreases \citep{zhu201830,
he2020mutinc}. In addition, $\sim 30$--$40\%$ of the SE systems are accompanied
by cold Jupiters \citep[CJs; masses $\gtrsim 0.5 M_{\rm J}$ and semi-major axes
$\gtrsim 1\;\mathrm{AU}$][]{zhu2018super, bryan2019excess} with
significantly inclined ($\gtrsim 10^\circ$) orbits relative to the SEs
\citep{masuda2020mutual}.

SEs are formed in gaseous protoplanetary disks, and likely have experienced an earlier
phase of giant impacts and collisions following the dispersal of disks
\citep[e.g.][]{liu2015giant, inamdar2016stealing, izidoro2017breaking}. As a
result, the SEs' initial obliquities are expected to be broadly distributed
\citep{li2020planetary, li2021giant}. However, due to the proximity of these
planets to their host stars, tidal dissipation can change the planets' spin
rates and orientations substantially within the age of the planetary system.
Indeed, the tidal spin-orbit alignment timescale is given by
\begin{align}
    t_{\rm al} \simeq{}& \p{30\;\mathrm{Myr}}
        \p{\frac{Q/2k_2}{10^3}}
        \p{\frac{M_\star}{M_{\odot}}}^{-3/2}\nonumber\\
    &\times \p{\frac{m}{4m_{\oplus}}}
        \p{\frac{R}{2R_{\oplus}}}^{-3}
        \p{\frac{a}{0.4\;\mathrm{AU}}}^{9/2},
        \label{eq:t_al}
\end{align}
where $m$, $R$, $k_2$, $Q$, and $a$ are the planet's mass, radius, tidal Love
number, tidal quality factor, and semi-major axis respectively, and $M_{\star}$
is the stellar mass. In a previous paper \citep{su2021dynamics}, we have studied
the combined effects of spin-orbit resonance and tidal dissipation in a
two-planet system (i.e.\ a SE with a companion), and showed that the planet's
spin can only evolve into to two possible long-term equilibria (``Tidal Cassini
Equilibria''), one of which can have a significant obliquity. In this paper, we
extend our analysis to three-planet systems consisting of either three SEs or
two SEs and a CJ\@. In addition to the equilibria analogous to those of the
two-planet case, we discover a novel class of oblique spin equilibria unique to
multi-planet systems. Such equilibria can substantially increase the occurrence
rate of oblique SEs in these architectures.

This paper is organized as follows. In Section~\ref{s:eom}, we summarize
the evolution of a SE in a multi-planetary system. In Section~\ref{s:align}, we
introduce a tidal alignment torque that damps the SE's obliquity
and investigate the resulting steady-state behavior. In
Section~\ref{s:weaktide}, we evolve both the SE spin rate and orientation
according to weak friction theory of the equilibrium tide. We show that the
qualitative dynamics are similar to the simpler model studied in
Section~\ref{s:align}. We summarize and discuss in Section~\ref{s:disc}.

\section{Spin Equations of Motion}\label{s:eom}

The unit spin vector $\uv{S}$ of an oblate planet orbiting a host star precesses
around the planet's unit angular momentum $\uv{L}$ following the equation
\begin{align}
    \rd{\uv{S}}{t}
        &= \alpha\p{\uv{S} \cdot \uv{L}}\p{\uv{S} \times \uv{L}}
        ,\label{eq:dsdt1}
\end{align}
where the characteristic spin-orbit precession frequency $\alpha$ is given by
\begin{align}
    \alpha ={}&
        \frac{3GJ_2 mR^2 M_\star}{2a^3 C\Omega_{\rm s}}
        = \frac{3k_q}{2k}\frac{M_\star}{m}\p{\frac{R}{a}}^3 \Omega_{\rm s}
            \nonumber\\
        ={}& \frac{1}{150\;\mathrm{yr}}
            \p{\frac{k_{\rm q}}{k}}
            \p{\frac{M_{\star}}{M_{\odot}}}^{3/2}\\
        &\times \p{\frac{m}{2M_{\oplus}}}^{-1}
            \p{\frac{R}{1.2R_{\oplus}}}^{3}
            \p{\frac{a}{0.1\;\mathrm{AU}}}^{-9/2}
            \frac{\Omega_{\rm s}}{n} \label{eq:wsl}.
\end{align}
In Eq.~\eqref{eq:wsl}, $\Omega_{\rm s}$ is the rotation rate of the planet,
$C = k mR^2$ is its moment of inertia (with $k$ the normalized moment of
inertia), $J_2 = k_{\rm q}\Omega_{\rm s}^2 (R^3/Gm)$ (with $k_{q}$ a constant)
is its rotation-induced (dimensionless) quadrupole moment, and $n \equiv
\sqrt{GM_\star / a^3}$ is its mean motion. For a SE, we adopt $k \sim k_{\rm q}
\sim 0.3$ \citep[e.g.][]{groten2004fundamental, lainey2016quantification}.

The orbital axis $\uv{L}$ also evolves in time, precessing and nutating about
the total angular momentum axis of the exoplanetary system, which we denote by
$\uv{J}$. When there are just two planets, this precession is uniform (with rate
$g$ and constant inclination angle between $\uv{L}$ and $\uv{J}$), and the spin
dynamics of the planet is described by the well-studied ``Colombo's Top''
system \citep{colombo1966, peale1969, peale1974possible, ward1975tidal,
henrard1987}. The spin equilibra of this system are termed ``Cassini States''
(CSs), and the number of CSs and their obliquities depend on the ratio $\eta
\equiv \abs{g} / \alpha$. In the presence of a tidal spin-orbit alignment
torque, up to two equilibria are stable and attracting, as shown in
[\citealp{su2021dynamics}; see also \citealp{fabrycky_otides, levrard2007,
peale2008obliquity}]: for $\eta \gg 1$, only CS2 is stable, with $\uv{S}$ nearly
aligned with $\uv{J}$; for $\eta \lesssim 1$, $\uv{S}$ can evolve towards two
possible states, the ``trivial'' CS1 with a small spin-orbit misalignment angle
$\theta_{\rm sl}$, or the ``resonant'' CS2 with significant $\theta_{\rm sl}$
(which approaches $90^\circ$ for $\eta \ll 1$).

When the SE is surrounded by multiple companions, the precession of $\uv{L}$
occurs on multiple characteristic frequencies \citep[see][]{murray1999solar}. In
this case, the spin dynamics given by Eq.~\eqref{eq:dsdt1} is complex and can
lead to chaotic behavior \citep[e.g.\ the chaotic obliquity evolution of Mars][]{
laskar1993chaotic, touma1993chaotic}. But what is the final equilibrium
state of $\uv{S}$ in the presence of tidal alignment? In this paper, we focus on
the case where the SE has two planetary companions. If the mutual inclinations
among the three planets are small, then the explicit solution for $\uv{L}(t)$ can
be written as \citep{murray1999solar}
\begin{align}
\mathcal{I} \equiv{}& I\exp\p{i\Omega}\nonumber\\
    ={}& I_{\rm (I)}\exp\s{ig_{\rm (I)}t + i\phi_{\rm (I)}}
        + I_{\rm (II)}\exp\s{ig_{\rm (II)}t + i\phi^{\rm
        (II)}},\label{eq:I_t_sol}\\
\uv{L} ={}& \Re\p{\mathcal{I}}\uv{X} + \Im\p{\mathcal{I}}\uv{Y}
        + \sqrt{1 - \abs{\mathcal{I}}^2}\,\uv{Z}.\label{eq:l_t_sol}
\end{align}
Here, $I$ is the inclination of $\uv{L}$ relative to $\uv{J}$, $\mathcal{I}$ is
the complex inclination, the quantities $I_{\rm (I, II)}$, $g_{\rm (I, II)}$
and $\phi_{\rm (I, II)}$ are the amplitudes, frequencies, and phase offsets of
the two inclination modes, indexed by (I) and (II), and the Cartesian coordinate
system $XYZ$ is defined such that $\uv{J} = \uv{Z}$. Without loss of generality,
we denote mode I as the dominant mode, with $I_{\rm (I)} \geq I_{\rm (II)}$. For
simplicity, we fix $\phi_{\rm (I)} = \phi_{\rm (II)} = 0$.

\section{Steady States Under Tidal Alignment Torque.}\label{s:align}

Since SEs are close to their host stars, tidal torques tend to drive $\uv{S}$
towards alignment with $\uv{L}$ and $\Omega_{\rm s}$ towards synchronization
with the mean motion (see Eq.~\ref{eq:t_al}). As the evolution of $\Omega_{\rm
s}$ also changes $\alpha$ (Eq.~\ref{eq:wsl}) and the underlying phase-space
structure, we first consider the dynamics when ignoring the spin magnitude
evolution. In this case, the planet's spin orientation experiences an alignment
torque, which we describe by
\begin{equation}
    \p{\rd{\uv{S}}{t}}_{\rm al}
        = \frac{1}{t_{\rm al}} \uv{S} \times \p{\uv{L} \times \uv{S}},
        \label{eq:dsdt_tide_toy}
\end{equation}
where $t_{\rm al}$ is given by Eq.~\eqref{eq:t_al}. Note that $t_{\rm al}$ is
significantly longer than all precession timescales in the system.

With two precessional modes for $\uv{L}(t)$, we expect that the tidally stable
spin equilibria (steady states) correspond to the stable, attracting CSs for
each mode, when they exist. In other words, if we denote the CS2 corresponding
to mode I by CS2(I), then we expect that the tidally stable equilibria are among
the four CSs: CS1(I) (if it exists), CS2(I), CS1(II) (if it exists), and
CS2(II). The corresponding CS obliquities $\theta_{\rm sl} \equiv
\cos^{-1}(\uv{S} \cdot \uv{L})$ are given by
\begin{equation}
    \alpha\cos \theta_{\rm sl} = - g_{\rm (I, II)}
        \p{\cos I_{\rm (I, II)} + \sin I_{\rm (I, II)}\cot \theta_{\rm sl}
            \cos \phi_{\rm sl}},
            \label{eq:def_csobl_th}
\end{equation}
where the azimuthal angle $\phi_{\rm sl}$ of $\uv{S}$ around $\uv{L}$ is
$\phi_{\rm sl} = 0$ (corresponding to $\uv{S}$ and $\uv{J}$ being coplanar but
on  opposite sides of $\uv{L}$) for CS1 and $\phi_{\rm sl} = \pi$ for CS2 [note
that because of the tidal alignment torque, the actual $\phi_{\rm sl}$ value is
slightly shifted from $0$ or $\pi$; see \citealp{su2021dynamics}].

\begin{figure*}
    \centering
    \includegraphics[width=1.5\columnwidth]{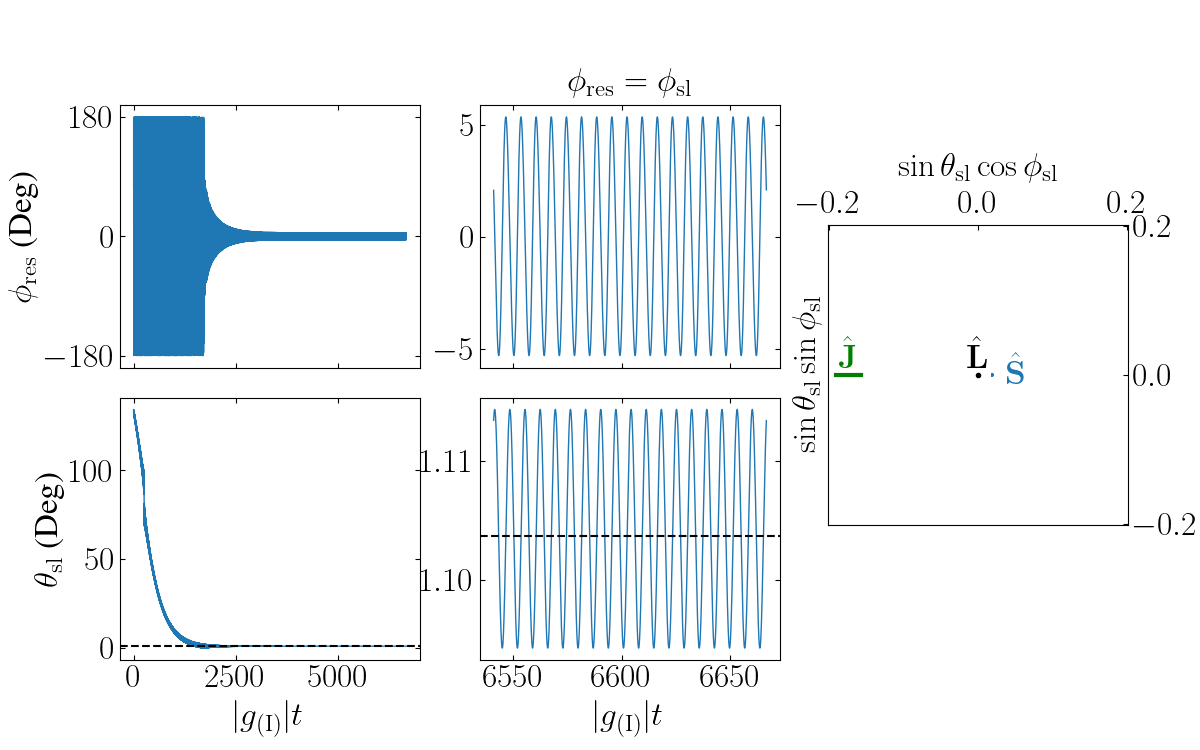}
    \includegraphics[width=1.5\columnwidth]{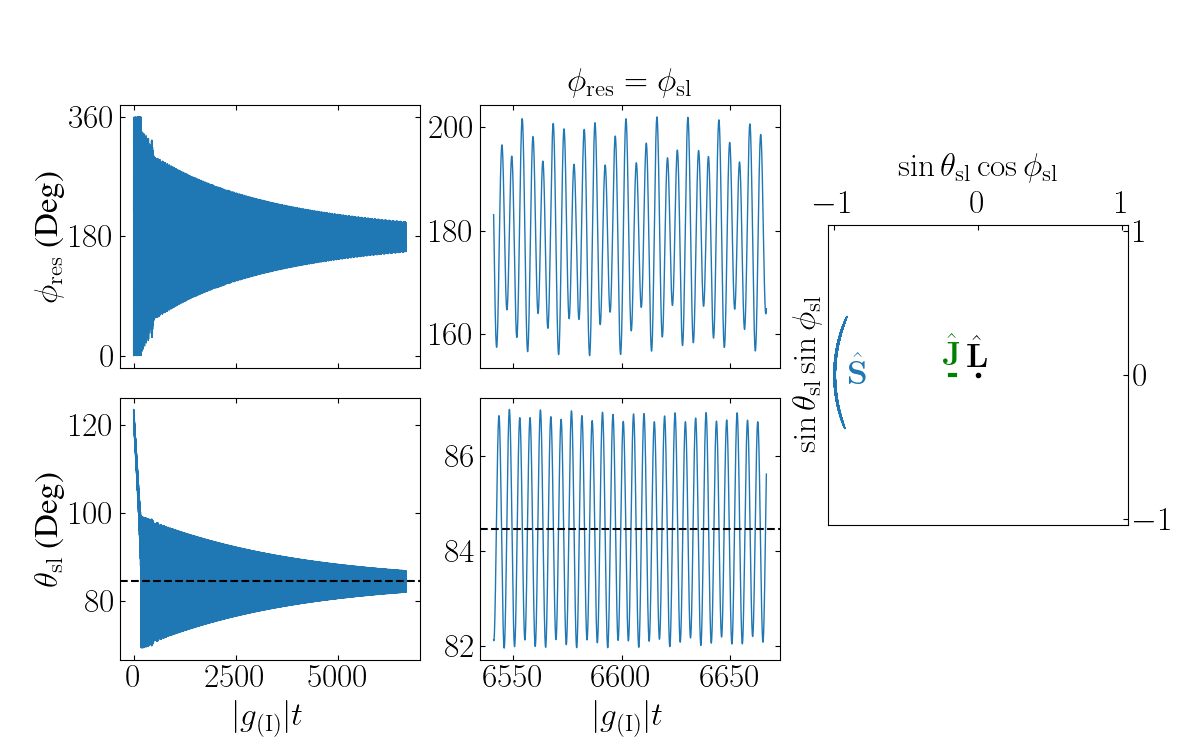}
    \caption{Two evolutionary trajectories of $\uv{S}$ showing capture into mode
    I resonances (CSs). In both cases, the mode parameters describing the
    evolution of $\uv{L}$ (see Eq.~\ref{eq:l_t_sol}) are $I_{\rm (I)} =
    10^\circ$, $I_{\rm (II)} = 1^\circ$, $g_{\rm (I)} = -0.1\alpha$, and $g_{\rm
    (II)} = 0.1g_{\rm (I)}$, while the initial spin orientations differ. In the
    top group of plots showing capture into CS1(I) [i.e.\ Cassini State 1 of
    mode I], the left four panels show the evolution of the spin obliquity
    $\theta_{\rm sl}$ and the resonant phase angle $\phi_{\rm res}$; in this
    case, $\phi_{\rm res}$ equals $\phi_{\rm sl}$, the azimuthal angle of
    $\uv{S}$ around $\uv{L}$, defined so that $\phi_{\rm sl} = 0$ corresponds to
    $\uv{S}$ and $\uv{J}$ being coplanar with $\uv{L}$ but on opposite sides of
    $\uv{L}$. The horizontal black dashed line shows the theoretically predicted
    obliquity of CS1(I), given by Eq.~\eqref{eq:def_csobl_th}. The right panel
    shows the final steady-state spin axis projected onto the orbital plane
    (perpendicular to $\uv{L}$). In these coordinates, $\uv{L}$ (black dot) is
    stationary, while $\uv{J}$ (green line) librates with a fixed orientation,
    and $\uv{S}$ is shown in blue. The bottom group of panels shows the same but
    for capture into the CS2(I) resonance.}\label{fig:example1_01}
\end{figure*}

To confront this expectation, we integrate
Eqs.~\eqref{eq:dsdt1},~\eqref{eq:l_t_sol}, and~\eqref{eq:dsdt_tide_toy}
numerically starting from various spin orientations.
Figure~\ref{fig:example1_01} shows two evolutionary trajectories for a system
with $I_{\rm (I)} = 10^\circ$, $I_{\rm (II)} = 1^\circ$, $\abs{g_{\rm (I)}} =
0.1\alpha$, and $\abs{g_{\rm (II)}} = 0.01\alpha$. Such a system can be
realized, for instance, by three SEs with masses $M_{\oplus}$, $3M_{\oplus}$,
and $3M_{\oplus}$ and semi-major axes $0.1\;\mathrm{AU}$, $0.15\;\mathrm{AU}$,
and $0.4\;\mathrm{AU}$ (see the left panels of
Fig.~\ref{fig:Ig_sweep} in the Appendix\footnote{Note that the mode amplitudes $I_{\rm (I)}$ and
$I_{\rm (II)}$ in Fig.~\ref{fig:Ig_sweep} are closer in magnitude than the case
we consider here. We exaggerate the inclination hierarchy for a more intuitive
physical picture, and explore the case where the modes are of comparable
amplitudes later in the paper and in Appendix~\ref{app:mixed_mode}.}). We see that
the initially retrograde spin is eventually captured into a steady state
centered around CS1 or CS2 of the dominant mode (i.e.\ mode I), with $\phi_{\rm
sl}$ librating around $0$ or $\pi$, respectively. The small oscillation of the
final $\theta_{\rm sl}$ is the result of perturbations from mode II\@.

\begin{figure*}
    \centering
    \includegraphics[width=1.3\columnwidth]{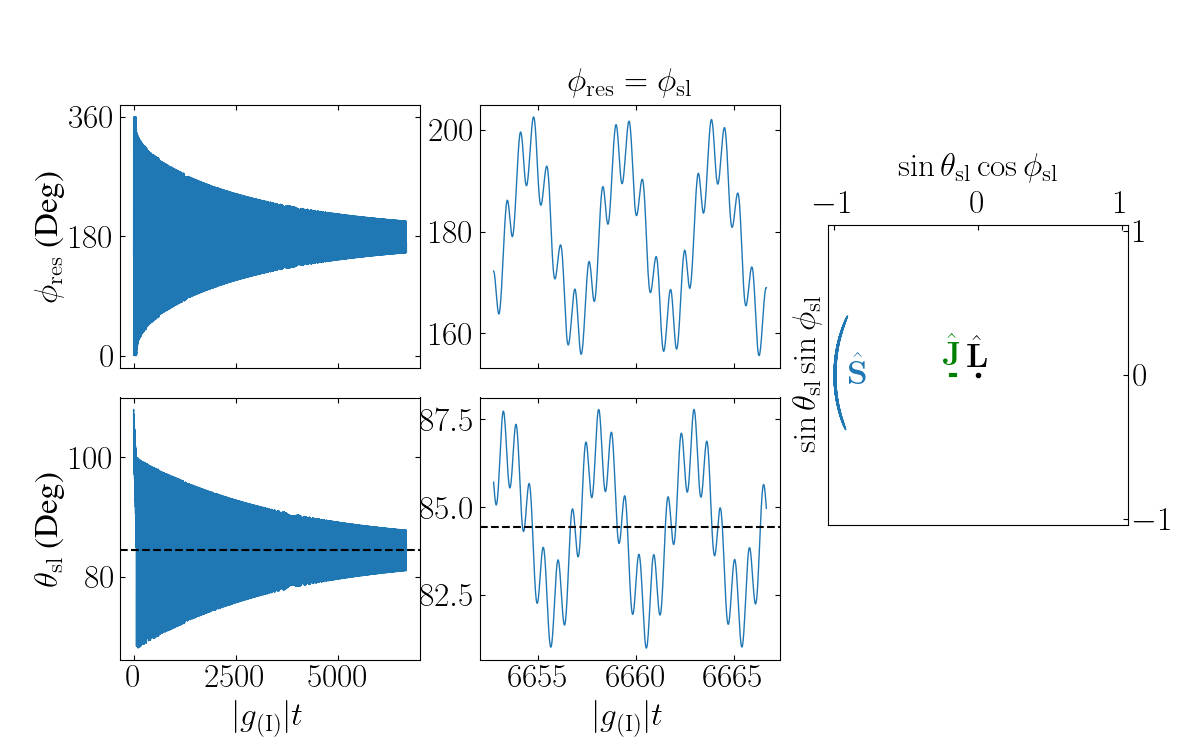}

    \includegraphics[width=1.3\columnwidth]{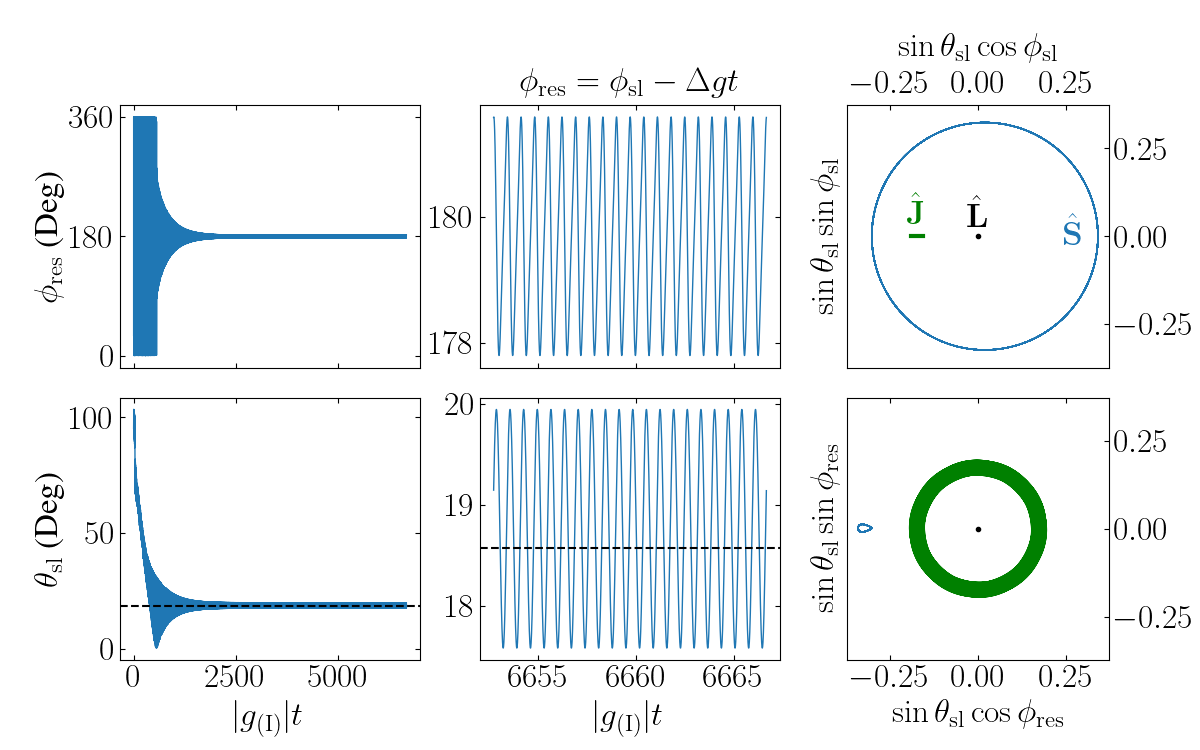}

    \includegraphics[width=1.3\columnwidth]{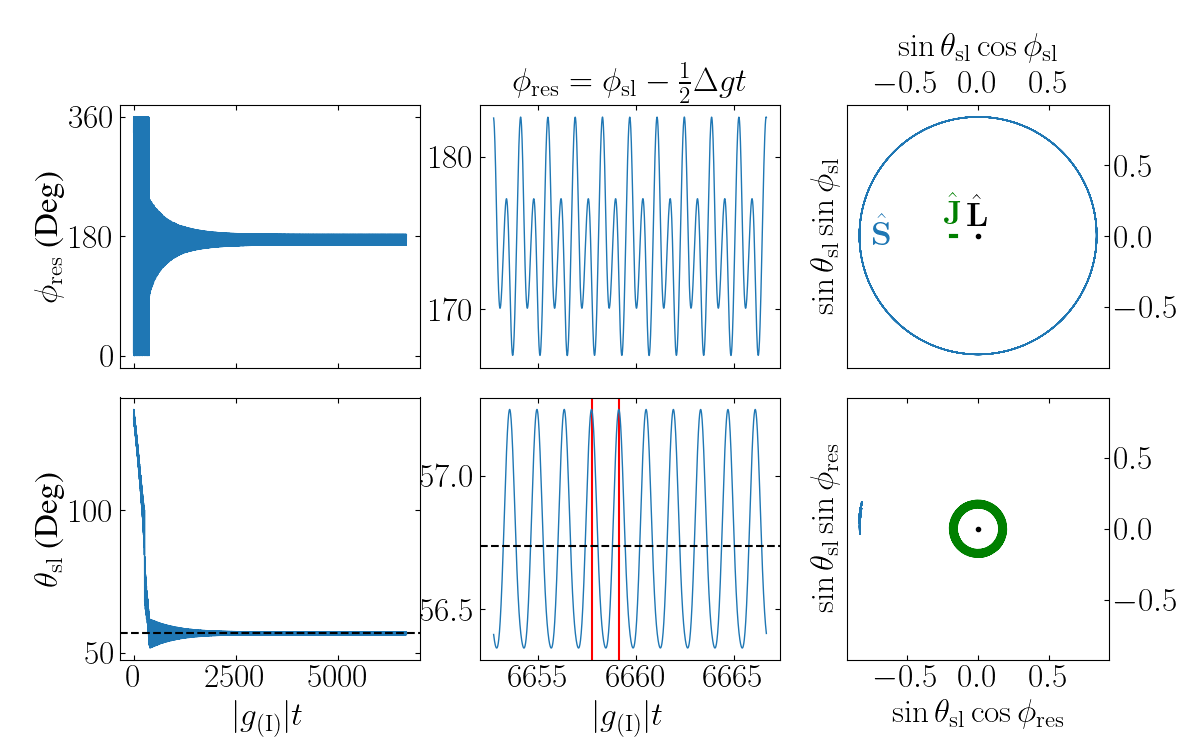}
    \caption{Three evolutionary trajectories for a system with the same
    parameters as in Fig.~\ref{fig:example1_01}, except for $g_{\rm (II)} =
    -\alpha = 10g_{\rm (I)}$. The three examples correspond to capture into
    CS2(I) (with $\phi_{\rm res} = \phi_{\rm sl}$), a resonance with $\phi_{\rm
    res} = \phi_{\rm sl} - \Delta g t$ [correpsonding to CS2(II)], and a ``mixed
    mode'' resonance with $\phi_{\rm res} = \phi_{\rm sl} - \Delta g t / 2$,
    where $\Delta g = g_{\rm (II)} - g_{\rm (I)}$. In the bottom two groups of
    plots, $\phi_{\rm sl}$ is not the resonant angle, so the spin axis encircles
    $\uv{L}$ in the top-right plots. For these two cases, we also display in the
    bottom-right panels the projection of the steady-state spin axis onto the
    $\uv{L}$ plane but with $\phi_{\rm res}$ as the azimuthal angle. In these
    two panels, $\uv{J}$ encircles $\uv{L}$ (as shown by the green ring), but
    the spin can be seen not to encircle $\uv{L}$, indicating that $\phi_{\rm
    res}$ is indeed librating. Finally, for the mixed-mode example (bottom
    group), the vertical red lines in the bottom-middle panel are separated by
    $2\pi / \abs{g_{\rm res}} = 4\pi / \abs{\Delta g}$, denoting the period of
    oscillation in $\theta_{\rm sl}$. }\label{fig:example1_10}
\end{figure*}

In addition to CS1(I) and CS2(I), we find that the spin can also settle down
into other equilibria (steady states) with different librating angles. In
general, we define the resonant phase angle
\begin{equation}
    \phi_{\rm res} \equiv \phi_{\rm sl} - g_{\rm res}t\label{eq:def_phires}.
\end{equation}
The examples shown in Fig.~\ref{fig:example1_01} correspond to $g_{\rm res} =
0$. In Fig.~\ref{fig:example1_10}, we show three evolutionary trajectories (with
three different initial spin orientations) of a system with the same parameters
as in Fig.~\ref{fig:example1_01} but with $g_{\rm (II)} = -\alpha$. Such a
system can be realized, for instance, by two warm SEs orbited by a cold Jupiter
(see the right panels of Fig.~\ref{fig:Ig_sweep25} in the Appendix where $a_3$
is small). Among these three examples, the first is captured into a resonance
with $g_{\rm res} = 0$ [i.e.\ CS2(I)], the second is captured into a resonance
with $g_{\rm res} = \Delta g \equiv g_{\rm (II)} - g_{\rm (I)}$, and the third
is captured into a resonance with $g_{\rm res} = \Delta g / 2$.

\begin{figure*}
    \centering
    \includegraphics[width=\columnwidth]{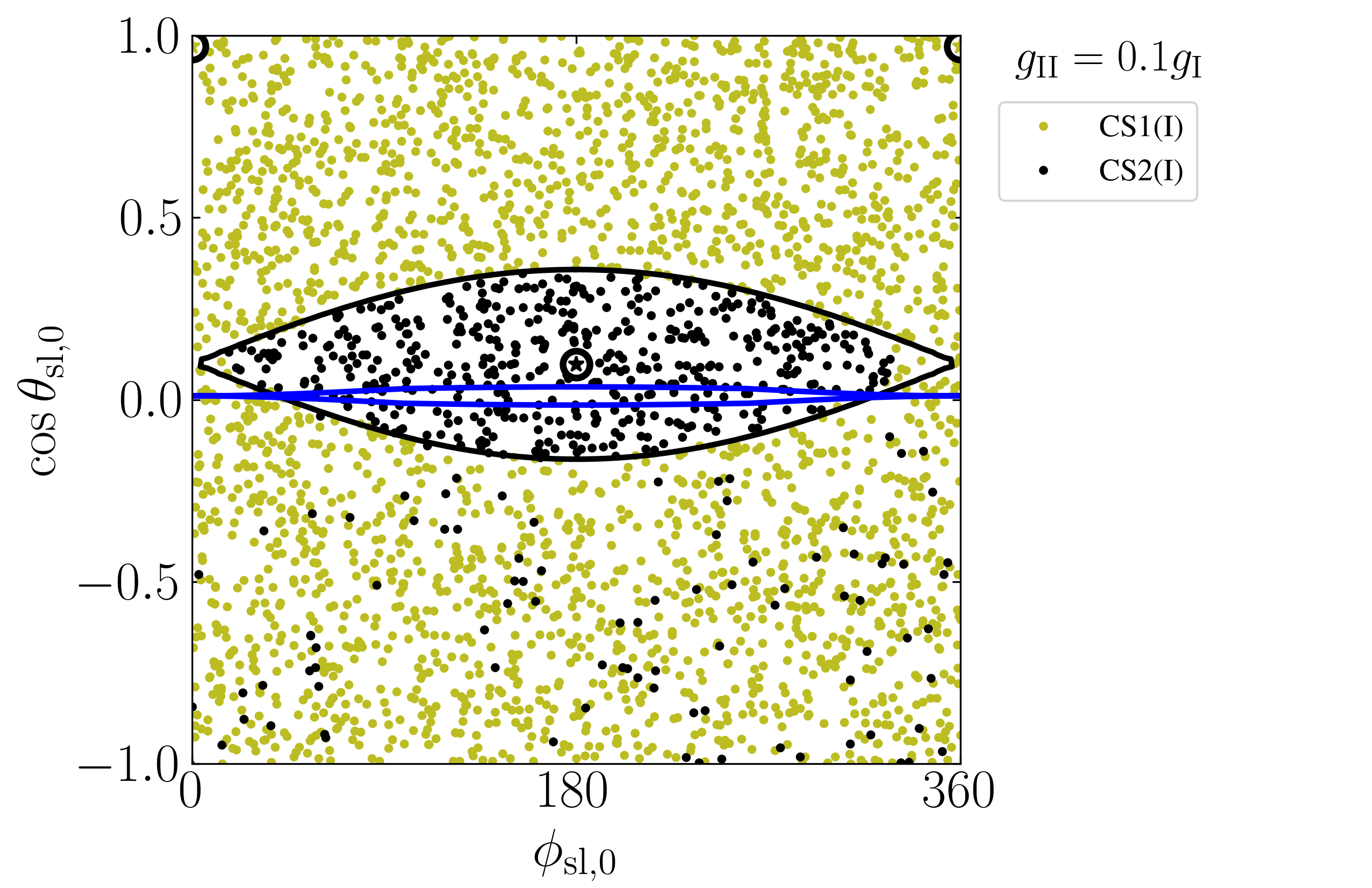}
    \includegraphics[width=\columnwidth]{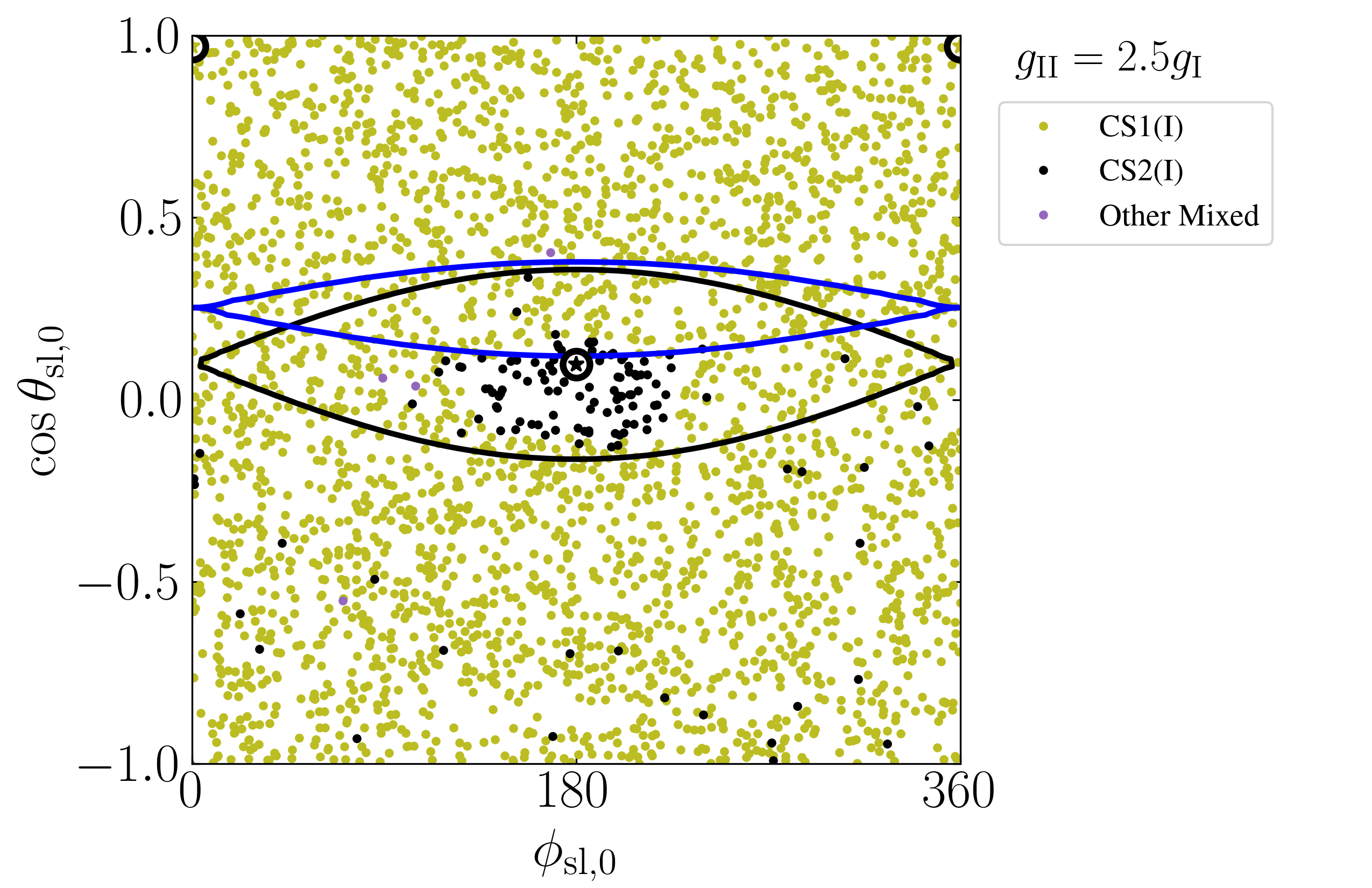}

    \includegraphics[width=\columnwidth]{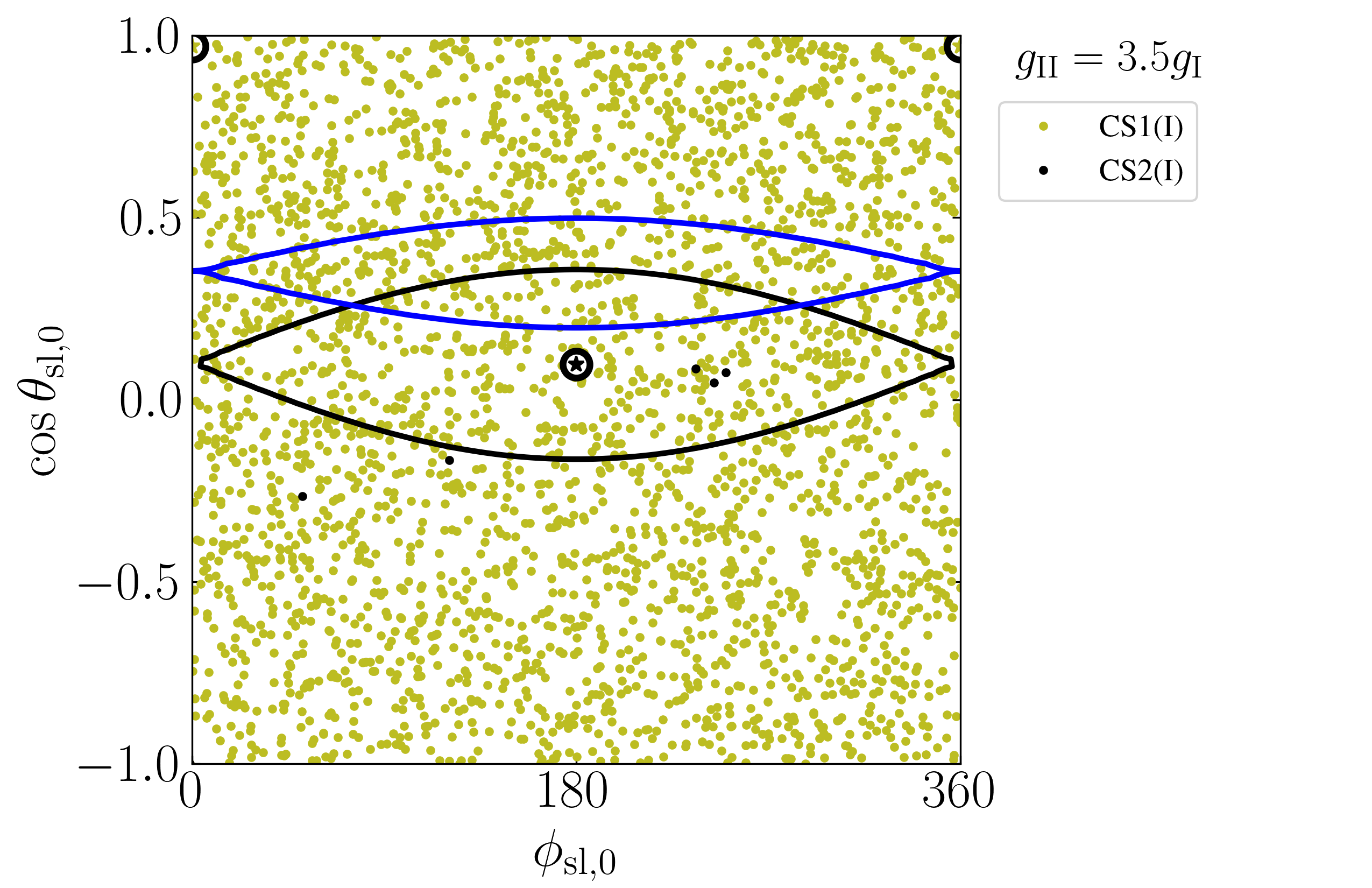}
    \includegraphics[width=\columnwidth]{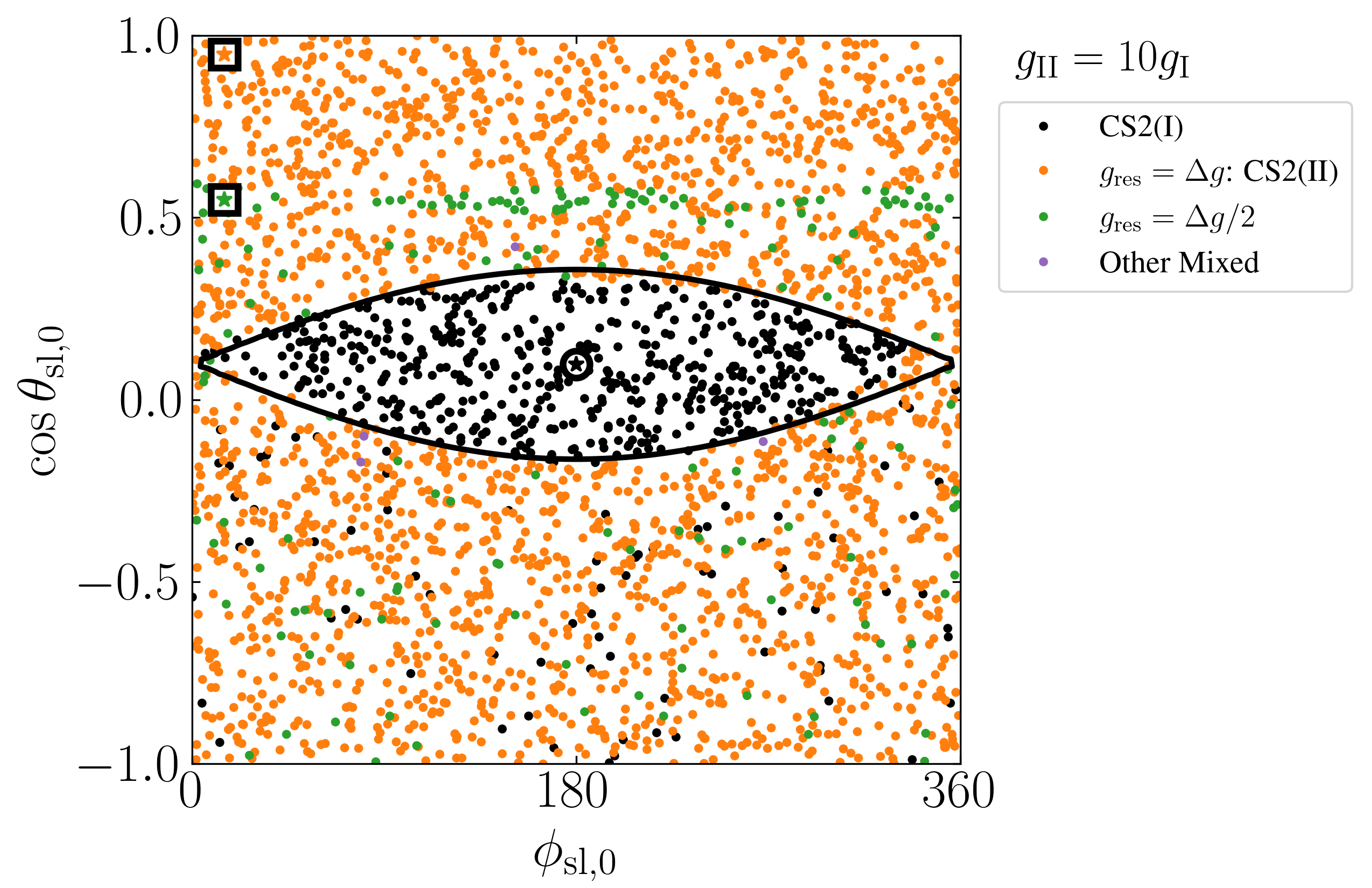}
    \caption{Asymptotic outcomes of spin evolution driven by tidal alignment
    torque for different initial spin orientations (described by $\theta_{\rm
    sl, 0}$ and $\phi_{\rm sl, 0}$) for a system with $I_{\rm (I)} = 10^\circ$,
    $I_{\rm (II)} = 1^\circ$, $\alpha = 10\abs{g_{\rm (I)}}$, and $g_{\rm (II)}
    = \z{0.1, 2.5, 3.5, 10} \times g_{\rm (I)}$ in the top-left, top-right,
    bottom-left, and bottom-right panels respectively (as labeled). Each dot
    represents an initial spin orientation, and the coloring of the dot
    indicates which Cassini State (CS) and which mode (legend) the system
    evolves into. The obliquity and $\phi_{\rm sl}$ values of the mode-I CSs are
    labeled as the circled stars, with the same colors as in the legend. The
    obliquities of the other equilibria are labeled as the boxed stars at the
    left edges of the right-bottom panel, with the same colors as in the legend;
    a small, arbitrary offset in $\phi_{\rm sl}$ is added to reflect the fact
    that these equilibria do not have fixed $\phi_{\rm sl}$ values. The
    separatrices for the mode I and mode II resonances are given in the
    black and blue lines respectively.}\label{fig:3outcomes1}
\end{figure*}

To explore the regimes under which various resonances are important, we
numerically determine [by integrating Eqs.~\eqref{eq:dsdt1},~\eqref{eq:l_t_sol},
and~\eqref{eq:dsdt_tide_toy}] the final spin equilibria (steady states) for
systems with different mode parameters ($I_{\rm (I)}$, $I_{\rm (II)}$, $g_{\rm
(I)}$, and $g_{\rm (II)}$), starting from all possible initial spin
orientations. Fig.~\ref{fig:3outcomes1} shows some examples of such a
calculation for $I_{\rm (I)} = 10^\circ$, $I_{\rm (II)} = 1^\circ$, $g_{\rm (I)}
= -0.1\alpha$ (the same as in
Figs.~\ref{fig:example1_01}--\ref{fig:example1_10}), but with the four different
values of $g_{\rm (II)} = \z{0.1, 2.5, 3.5, 10} \times g_{\rm (I)}$. We identify
three qualitatively different regimes:
\begin{itemize}
    \item When $\abs{g_{\rm (II)}} \ll \abs{g_{\rm (I)}}$ (top-left panel of
        Fig.~\ref{fig:3outcomes1}), mode II serves as a slow, small-amplitude
        perturbation to the dominant mode I, and the spin evolution is very
        similar to that studied in \citet{su2021dynamics}: prograde initial
        conditions (ICs) outside of the mode-I resonance evolve towards CS2(I),
        ICs inside the resonance evolve to CS2(I), and retrograde ICs outside of
        the mode-I resonance reach one of the two probabilistically.

    \item When $\abs{g_{\rm (II)}} \sim \abs{g_{\rm (I)}}$ (see the top-right
        and bottom-left panels of Fig.~\ref{fig:3outcomes1}), the resonances
        corresponding to the two modes overlap, chaotic obliquity evolution
        occurs \citep[see][]{touma1993chaotic, laskar1993chaotic}, and we expect
        that CS2(I) becomes less stable\footnote{An more precise resonance
        overlap condition can be obtained by comparing the separatrix widths, as
        the mode-II resonance is much narrower even when $g_{\rm (I)} = g_{\rm
        (II)}$. Such a condition would require $g_{\rm (I)}\sin I_{\rm (I)} \sim
        g_{\rm (II)} \sin I_{\rm (II)}$. Thus, the onset of chaos due to
        resonance overlap occurs somewhere in the range $I_{\rm (II)} / I_{\rm
        (I)} \lesssim g_{\rm (II)} / g_{\rm (I)} \lesssim 1$.}. Indeed we see
        that in this regime, fewer ICs evolve into the high-obliquity CS2(I)
        equilibrium of the dominant mode I, and most ICs lead to the
        low-obliquity CS1(I).

    \item When $\abs{g_{\rm (II)}} \gg \abs{g_{\rm (I)}}$ (see the bottom-right
        panel of Fig.~\ref{fig:3outcomes1}), the separatrix for mode II does not
        exist, we see that all ICs inside the separatrix of mode I again
        converge successfully to CS2(I), and CS2(II) becomes the preferred
        low-obliquity equilibrium. Additionally, a narrow band of ICs with $\cos
        \theta_{\rm sl, 0} \sim 0.6$ and some other scattered ICs with $\cos
        \theta_{\rm sl, 0} \lesssim 0.6$ evolve to the mixed-mode equilibrium
        with $g_{\rm res} = \Delta g / 2$, which has $\theta_{\rm eq} \approx
        60^\circ$. A second mixed-mode resonance with $g_{\rm res} = \Delta g /
        3$ is also observed (with $\theta_{\rm eq} \approx 67^\circ$) for a
        sparse set of ICs.
\end{itemize}

\begin{figure*}
    \centering
    \includegraphics[width=1.5\columnwidth]{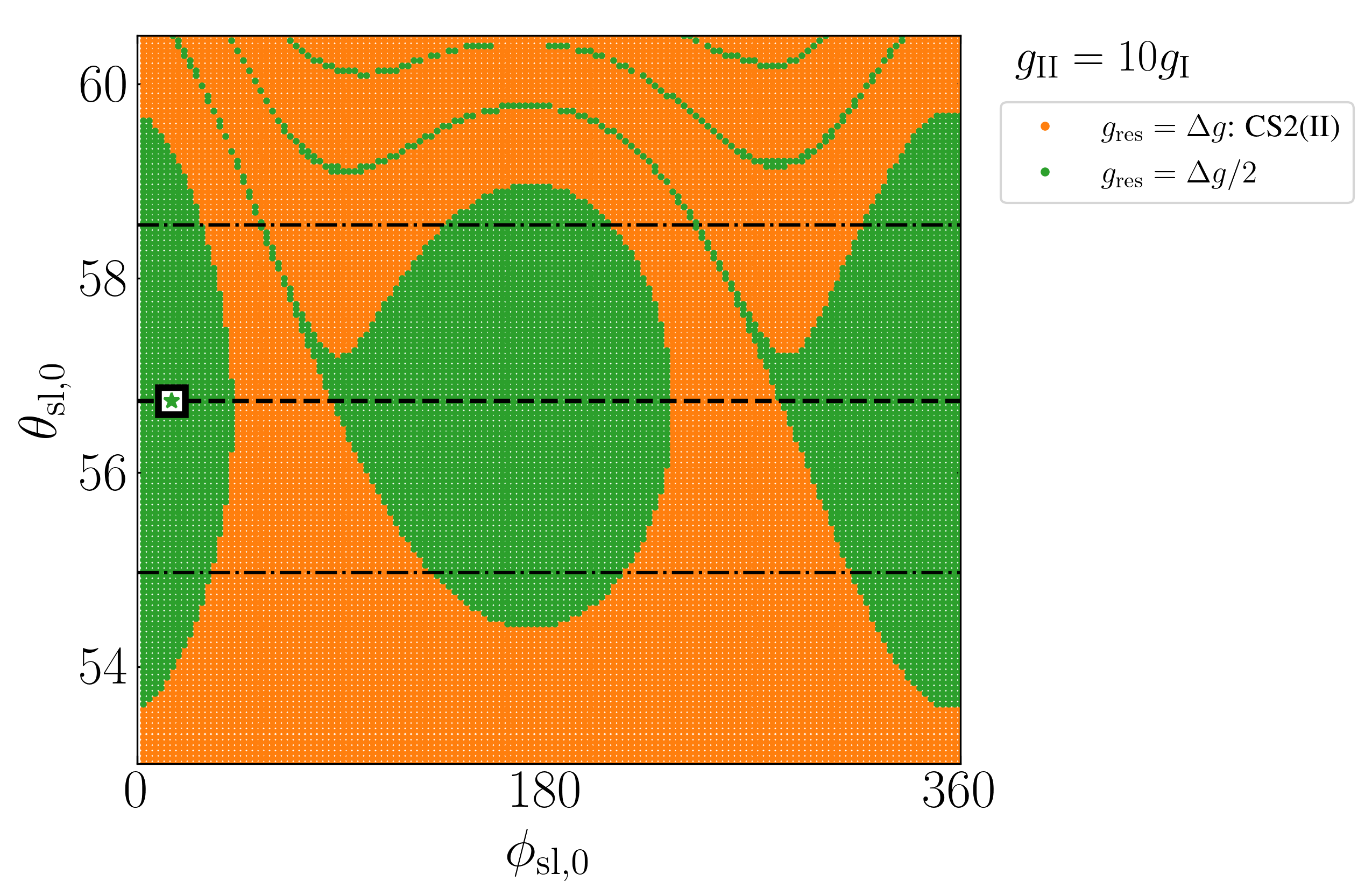}
    \caption{Same as the bottom right panel of Fig.~\ref{fig:3outcomes1} but
    zoomed-in to a narrow range of initial obliquities near $\theta_{\rm eq}
    \approx 57^\circ$ (horizontal dashed line) for the $g_{\rm res} = \Delta g /
    2$ mixed mode equilibrium as predicted by Eq.~\eqref{eq:g_res_rel}. The
    horizontal dash-dotted lines indicate the amplitude of oscillation of the
    mode (see Fig.~\ref{fig:example1_10}). It is clear that there are two basins
    of attraction for this mixed-mode resonance near $\phi_{\rm sl, 0} = 0$ and
    $\phi_{\rm sl, 0} = 180^\circ$.
    }\label{fig:outcomes_grid}
\end{figure*}

Towards a better understanding of how systems are captured into these mixed-mode
equilibria, we numerically calculate the ``basin of attraction'' by repeating
the procedure for producing Fig.~\ref{fig:3outcomes1} but instead use a fine
grid of ICs with $\theta_{\rm sl, 0}$ near the average obliquity of the
equilibrium. This results in a ``zoomed-in'' version of the bottom-right panel
of in Fig.~\ref{fig:3outcomes1} and doubles as a numerical stability analysis of
the equilibrium. Figure~\ref{fig:outcomes_grid} shows the result of this
procedure applied to the $g_{\rm res} = \Delta g / 2$ resonance, where we have
zoomed in to $\theta_{\rm sl, 0}$ near the $\theta_{\rm eq} \approx 60^\circ$
associated with the resonance. We see that the resonance is reached consistently
from some well-defined regions in the $\p{\theta_{\rm sl}, \phi_{\rm sl}}$
space.

\begin{figure}
    \centering
    \includegraphics[width=\columnwidth]{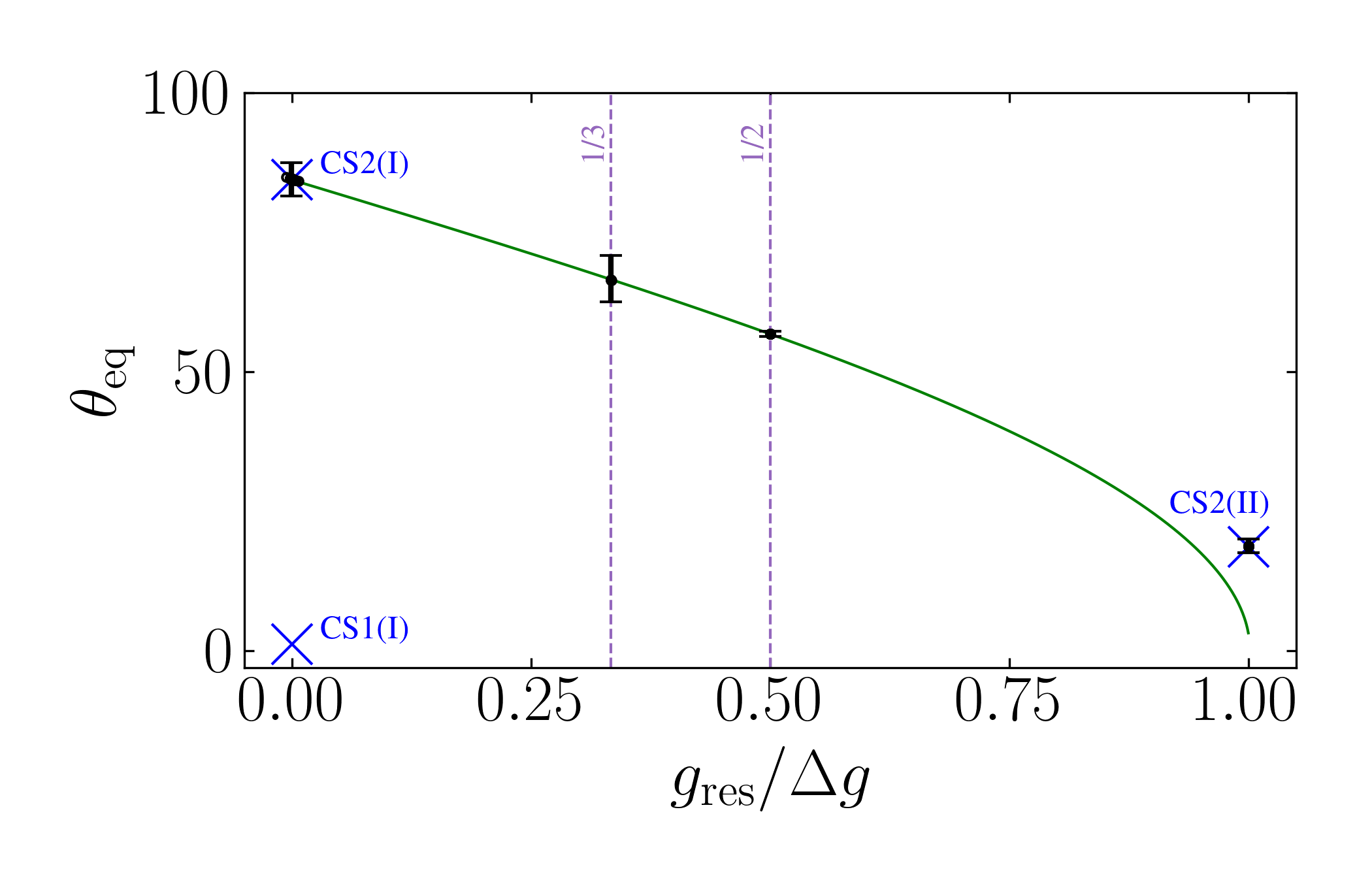}
    \caption{``Equilibrium'' obliquity as a function of the resonant frequency
    (Eq.~\ref{eq:def_phires}) for a system with $I_{\rm (I)} = 10^\circ$,
    $I_{\rm (II)} = 1^\circ$, $\alpha = 10\abs{g_{\rm (I)}}$, and $g_{\rm (II)}
    = 10g_{\rm (I)}$ (corresponding to the bottom-right panel of
    Fig.~\ref{fig:3outcomes1}). The green line is the analytical result given by
    Eq.~\eqref{eq:g_res_rel}. The error bars denote the amplitude of oscillation
    of $\theta_{\rm sl}$ when the spin has reached a steady state (e.g.\
    Figs.~\ref{fig:example1_01}--\ref{fig:example1_10}). The blue crosses denote
    the CSs for the two inclination modes. }\label{fig:outcomes10_scat}
\end{figure}

Figure~\ref{fig:outcomes10_scat} summarizes the equilibrium obliquities
$\theta_{\rm eq}$ of various resonances achieved for the system depicted in the
bottom-right panel of Fig.~\ref{fig:3outcomes1}. For a trajectory that reaches a
particular equilibrium, we compute $\theta_{\rm eq} =\ev{\theta_{\rm sl}}$ by
averaging over the last $100 / \max\p{\abs{g_{\rm (I)}}, \abs{\Delta g}}$ of the
integration. We obtain the corresponding ``resonance'' frequency by $g_{\rm res}
\simeq \ev{\dot{\phi}_{\rm sl}}$.

In fact, the relation between $\theta_{\rm eq}$ and $g_{\rm res}$ can be
described analytically. We consider the equation of motion in the rotating frame
where $\uv{L} = \uv{z}$ is constant and $\uv{J}$ lies in the $xz$ plane (i.e.\
$\uv{J} = -\sin I \uv{x} + \cos I \uv{z}$):
\begin{align}
    \p{\rd{\uv{S}}{t}}_{\rm rot} &= \alpha\p{\uv{S} \cdot
        \uv{L}}\p{\uv{S} \times \uv{L}}
        + \uv{S} \times \p{\dot{\Omega}\uv{J} +
        \dot{I}\uv{y}}.\label{eq:app_dsdtrot}
\end{align}
Let $\uv{S} = \sin \theta_{\rm sl}\p{\cos \phi_{\rm sl}\uv{x} + \sin \phi_{\rm
sl} \uv{y}} + \cos \theta_{\rm sl} \uv{z}$. The evolution of $\phi_{\rm sl}$
then follows
\begin{align}
    \rd{\phi_{\rm sl}}{t} ={}& -\alpha \cos\theta_{\rm sl}
        - \dot{\Omega}\p{\cos I + \sin I \cot \theta_{\rm sl} \cos \phi_{\rm
            sl}}\nonumber\\
        &- \dot{I} \cot \theta_{\rm sl} \sin \phi_{\rm sl}.
        \label{eq:app_dfdt}
\end{align}
Note that the single-mode CSs satisfy Eq.~\eqref{eq:app_dfdt} where
$\dot{\Omega} = g$, $\dot{I} = 0$, and $\phi_{\rm sl}$ is either equal to
$0^\circ$ or $180^\circ$ (Eq.~\ref{eq:def_csobl_th}). For the general, two-mode
problem, if $\phi_{\rm res} = \phi_{\rm sl} - g_{\rm res}t$ is a resonant angle,
then it must satisfy
\begin{equation}
    \ev{\rd{\phi_{\rm res}}{t}} = \ev{\rd{\phi_{\rm sl}}{t}} - g_{\rm res} = 0,
        \label{eq:constr}
\end{equation}
where the angle brackets denote an average over a libration period. Since
$\phi_{\rm sl}$ circulates when $g_{\rm res} \neq 0$, $\ev{\cos \phi_{\rm
sl}} = \ev{\sin \phi_{\rm sl}} = 0$. Furthermore, if $I_{\rm (II)} / I_{\rm
(I)}\equiv \epsilon \ll 1$, we can expand Eq.~\eqref{eq:I_t_sol} to obtain
\begin{align}
    \dot{\Omega} &= g_{\rm (I)}
        + \Delta g \epsilon \cos\p{\Delta g t}
        + \mathcal{O}\s{\epsilon^2},\label{eq:app_Wdot}\\
    I &= I_{\rm (I)}\p{
        1 + \epsilon \cos\p{\Delta g t}
        + \mathcal{O}\s{\epsilon^2}},\label{eq:app_Idot}
\end{align}
where $\Delta g \equiv g_{\rm (II)} - g_{\rm (I)}$. To leading order, we have
$\dot{\Omega} \simeq g_{\rm (I)}$ and $I \simeq I_{\rm (I)}$, so
Eq.~\eqref{eq:app_dfdt} reduces to
\begin{equation}
    \alpha \cos \theta_{\rm eq} \simeq
        - g_{\rm (I)}\cos I_{\rm (I)}
        - g_{\rm res}.\label{eq:g_res_rel}
\end{equation}
This is Eq.~\eqref{eq:g_res_rel} in the main text, and is shown in
Fig.~\ref{fig:outcomes10_scat}. Good agreement between the analytic expression
and numerical results is observed. Note that setting $g_{\rm res} = \Delta g$ in
Eq.~\eqref{eq:g_res_rel} does not yield the mode II CSs, as the mode I
inclination is still being used, and the $\phi_{\rm sl}$ terms are averaged out
in the mixed mode calculation while being nonzero in the CS obliquity
calculation.

\begin{figure*}
    \centering
    \includegraphics[width=0.8\textwidth]{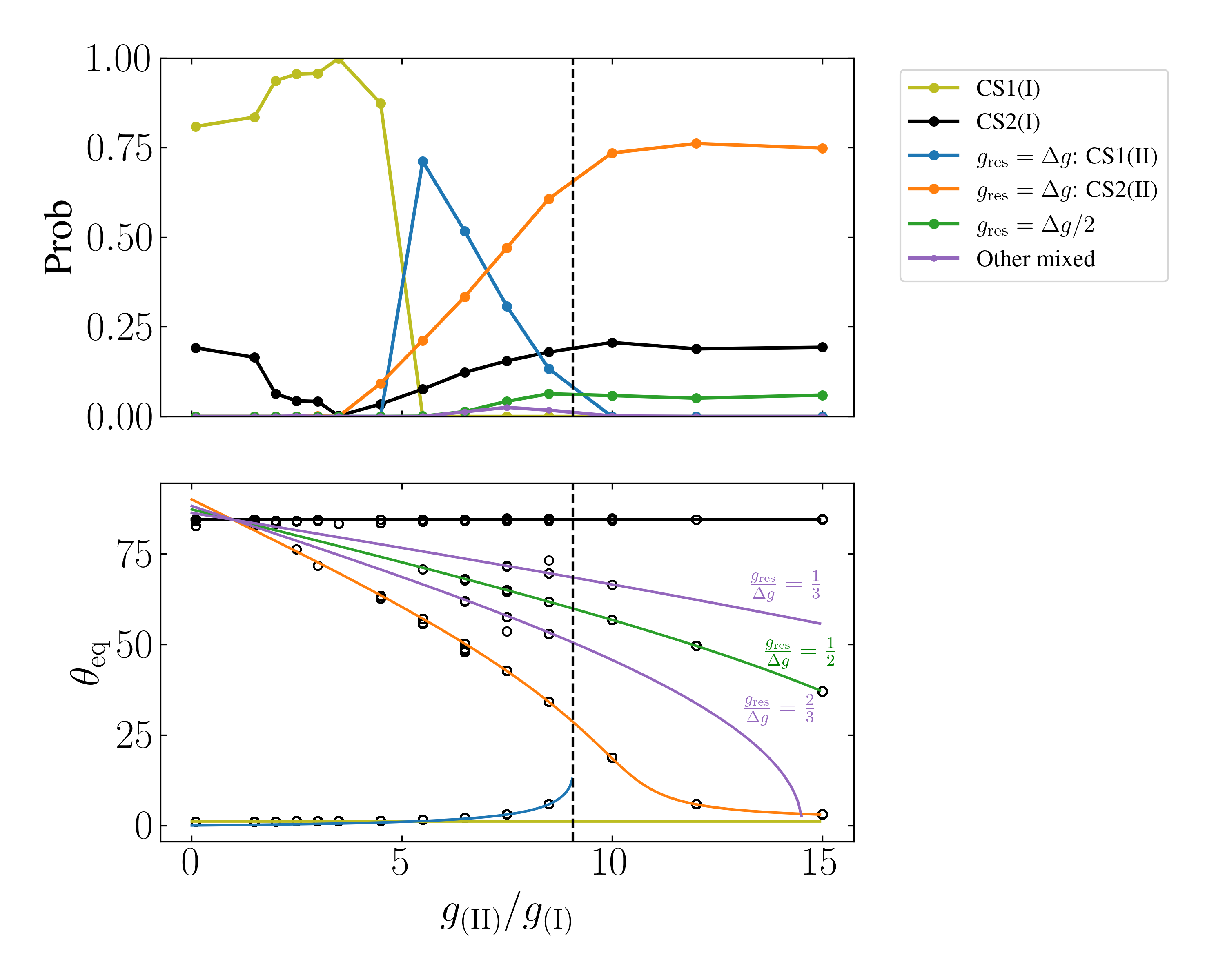}
    \caption{Final outcomes of spin evolution under tidal alignment torque for a
    3-planet system with inclination mode parameters $I_{\rm (I)} = 10^\circ$,
    $I_{\rm (II)} = 1^\circ$, $\alpha = 10\abs{g_{\rm (I)}}$ (same as
    Figs.~\ref{fig:example1_01}--\ref{fig:outcomes10_scat}) and varying $g_{\rm
    (II)} / g_{\rm (I)}$. The top panel shows the probability of each of the
    possible steady-state outcomes for $3000$ initial spin orientations sampled
    from an isotropic distribution. The vertical dashed line shows the value of
    $\abs{g_{\rm (II)}}$ above which CS1(II) no longer exists. The bottom panel
    shows the final equilibrium obliquities (open black circles) for each
    $g_{\rm (II)} / g_{\rm (I)}$. For the mixed-mode resonances ($g_{\rm res}
    \neq 0, \Delta g$), the equilibrium obliquities are given by
    Eq.~\eqref{eq:g_res_rel} and are shown as the solid green and purple lines
    for the labeled values of $g_{\rm res}$. The other lines are the equilibrium
    $\theta_{\rm eq}$ for ``pure'' CSs (as labeled), whose equilibria satisfy
    Eq.~\eqref{eq:def_csobl_th}. Not all observed mixed-mode resonances are
    plotted (e.g.\ for $g_{\rm (II)} = 7.5g_{\rm (I)}$, there is an outcome with
    $g_{\rm res} / \Delta g = 3/4$).
    }\label{fig:outcomes1}
\end{figure*}

The four systems shown in Fig.~\ref{fig:3outcomes1} demonstrate that the
characteristic spin evolution depends strongly on the ratio $g_{\rm (II)} /
g_{\rm (I)}$. To understand the transition between these different regimes, we
vary $g_{\rm (II)}$ over a wide range of values (while keeping the other
parameters the same as in Fig.~\ref{fig:3outcomes1}). For each $g_{\rm (II)}$,
we numerically determine the steady-state (equilibrium) obliquities and compute
the probability of reaching each equilibrium by evolving $3000$ initial spin
orientations drawn randomly from an isotropic distribution.
Figure~\ref{fig:outcomes1} shows the result. Two trends can be seen: the
probability of long-lived capture into the CS2(I) resonance decreases as
$\abs{g_{\rm (II)}}$ is increased from $\abs{g_{\rm (II)}} \ll \abs{g_{\rm
(I)}}$ to $\abs{g_{\rm (II)}} \sim \abs{g_{\rm (I)}}$, and mixed-mode resonances
become significant, though non-dominant, outcomes for $\abs{g_{\rm (II)}} \gg
\abs{g_{\rm (I)}}$.

\begin{figure*}
    \centering
    \includegraphics[width=\columnwidth]{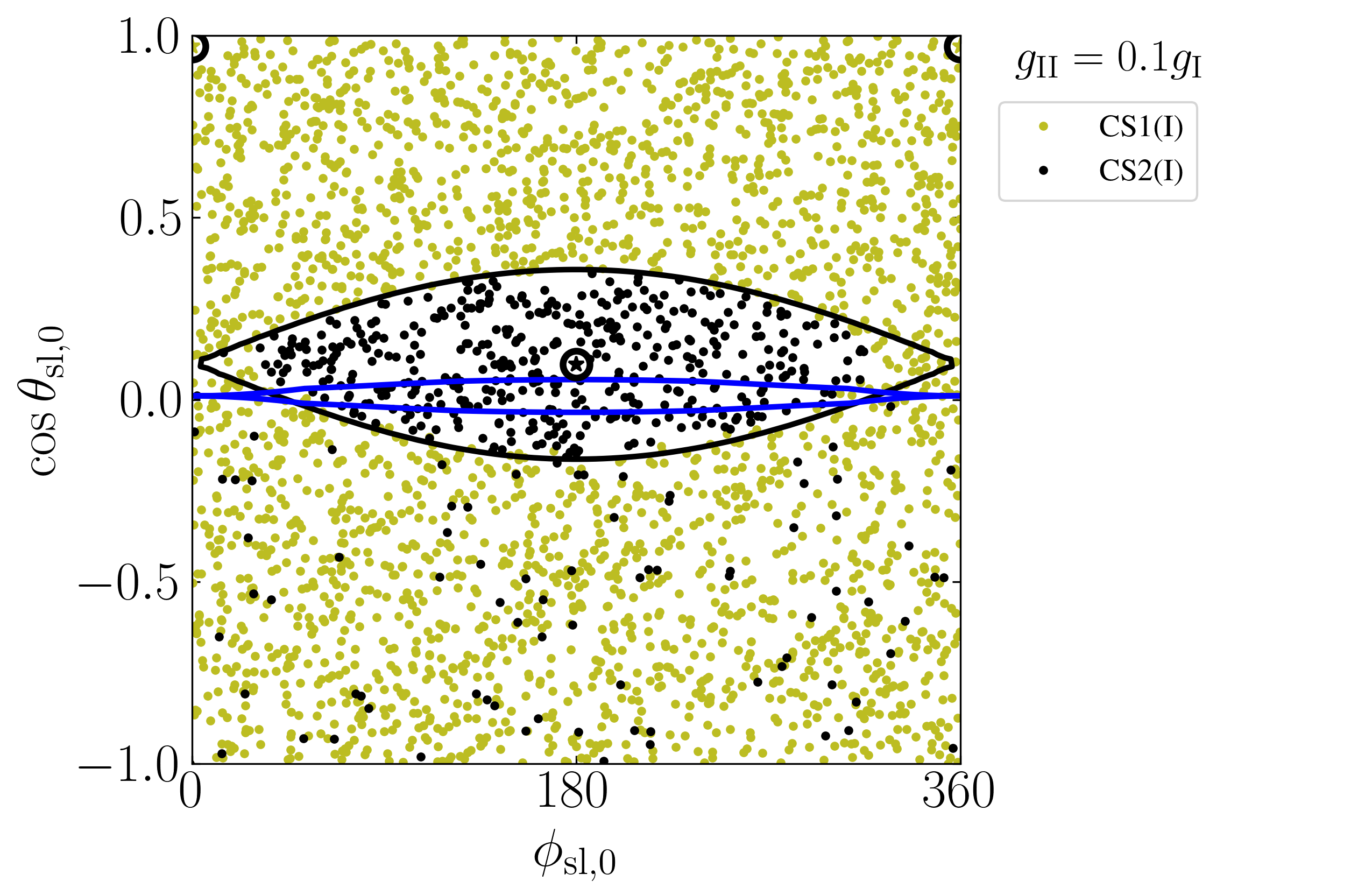}
    \includegraphics[width=\columnwidth]{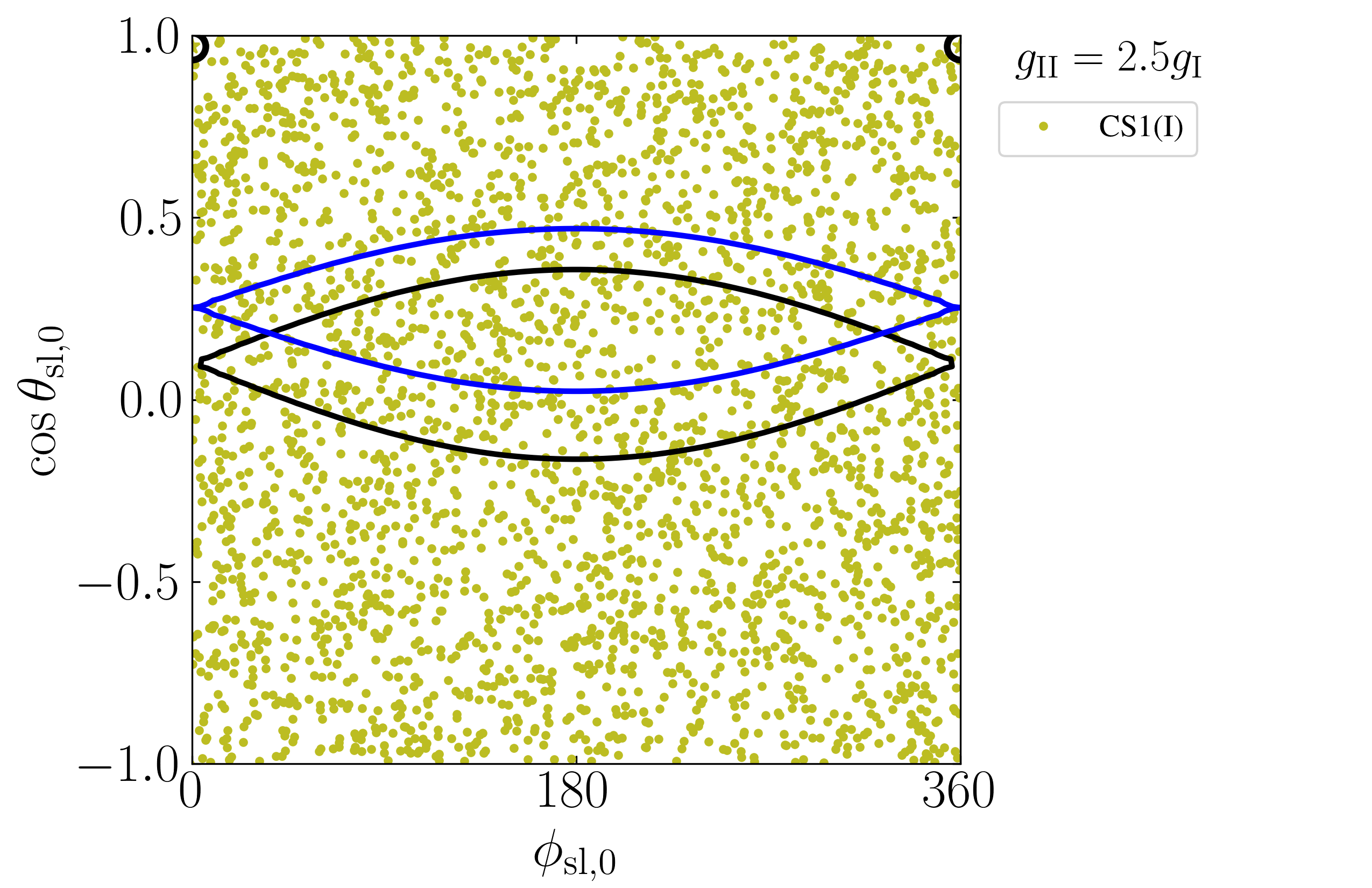}
    \includegraphics[width=\columnwidth]{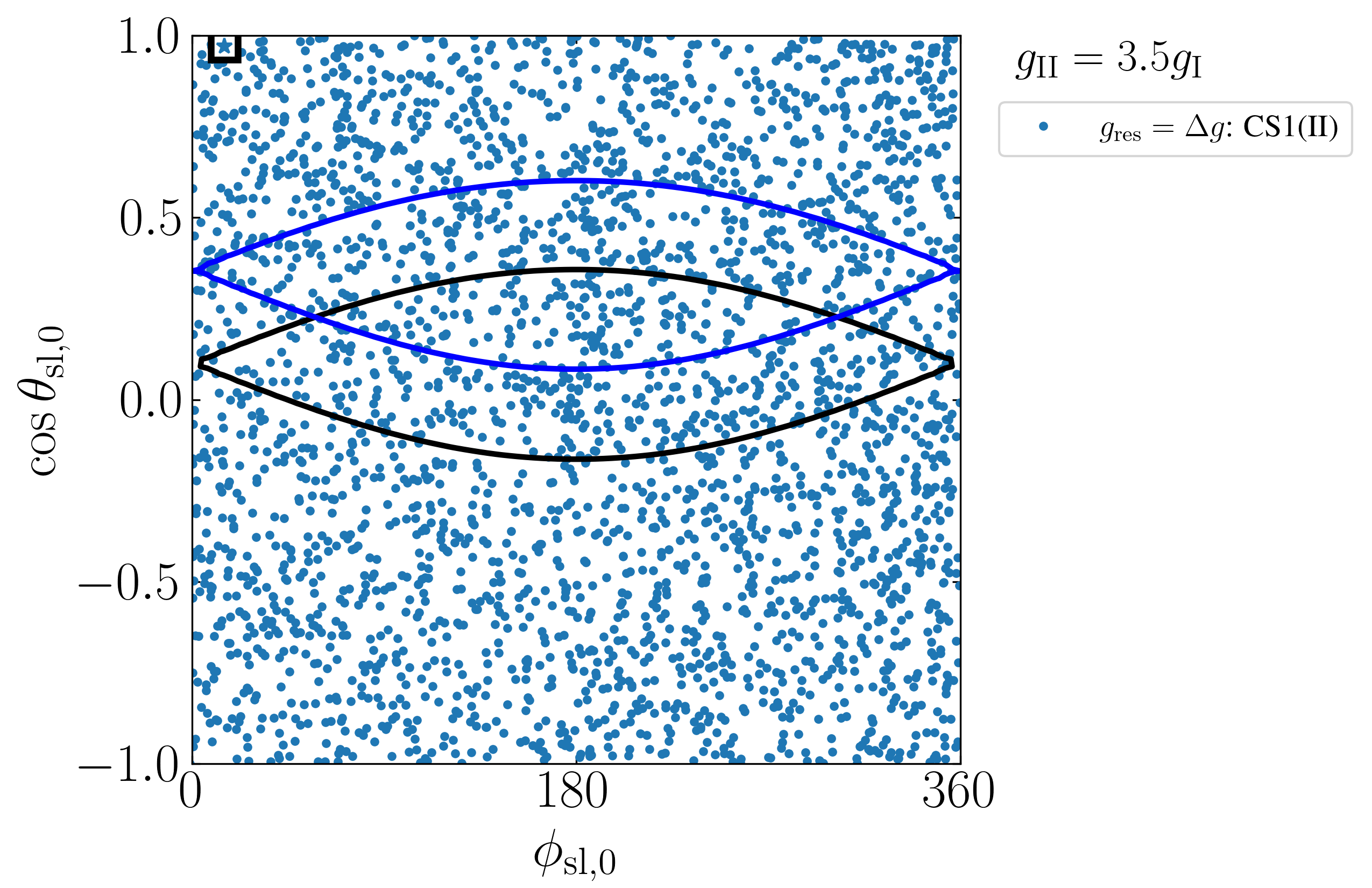}
    \includegraphics[width=\columnwidth]{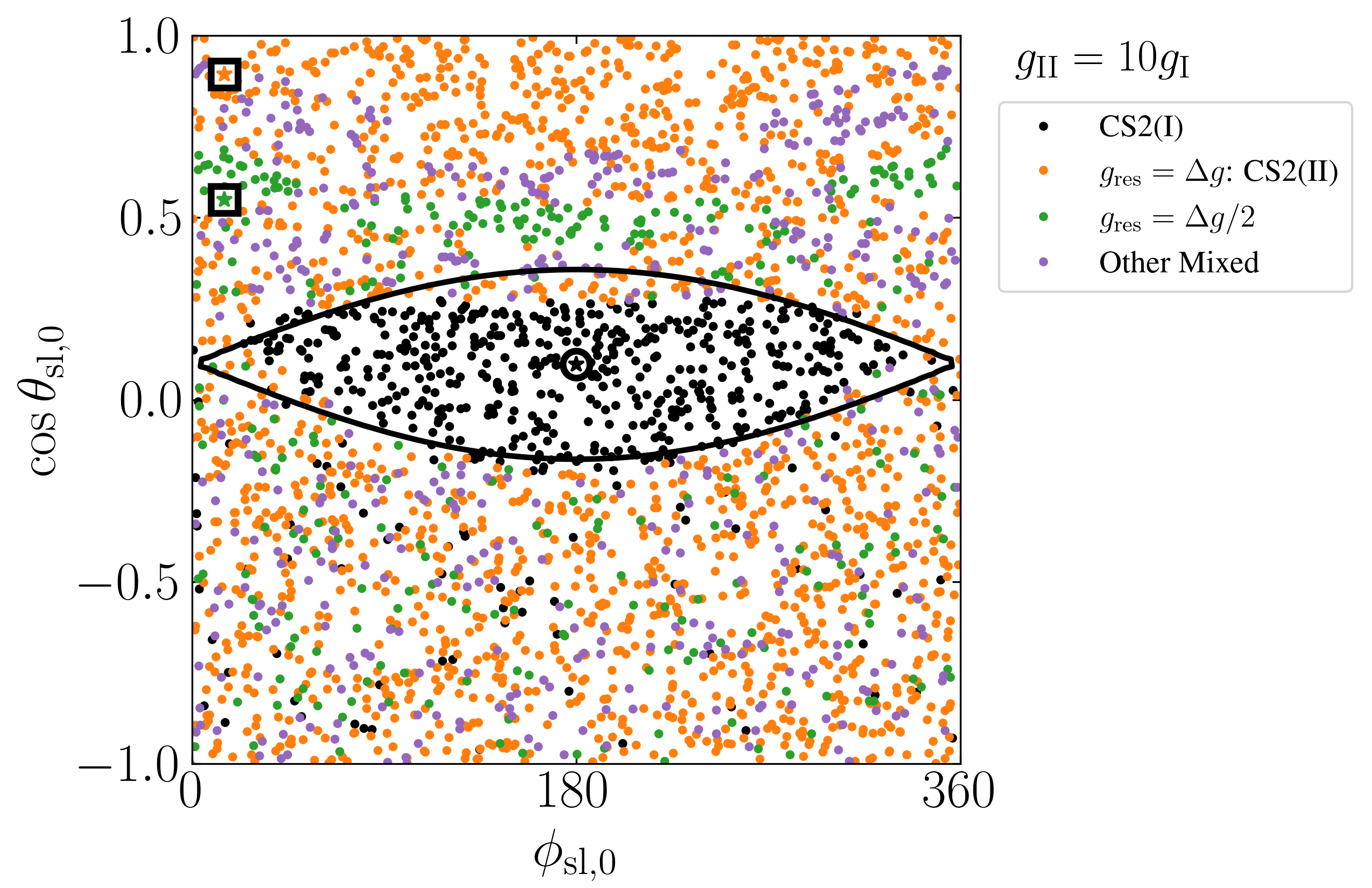}
    \caption{Same as Fig.~\ref{fig:3outcomes1} but for $I_{\rm (II)} = 3^\circ$.
    }\label{fig:3outcomes3}
\end{figure*}

\begin{figure}
    \centering
    \includegraphics[width=\columnwidth]{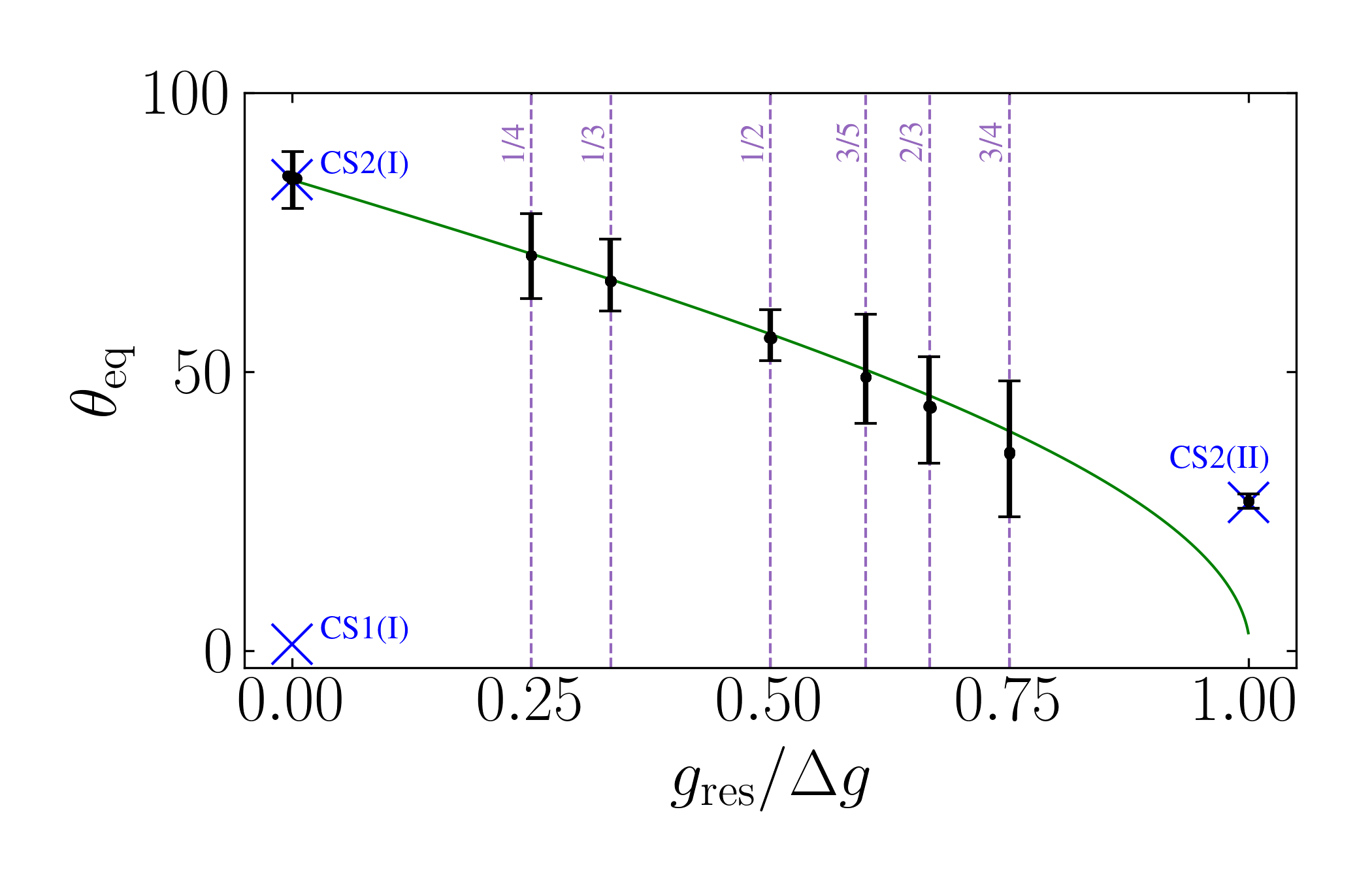}
    \caption{Same as Fig.~\ref{fig:outcomes10_scat} but for $I_{\rm (II)} =
    3^\circ$.}\label{fig:3outcomes010_scat}
\end{figure}

\begin{figure*}
    \centering
    \includegraphics[width=0.75\textwidth]{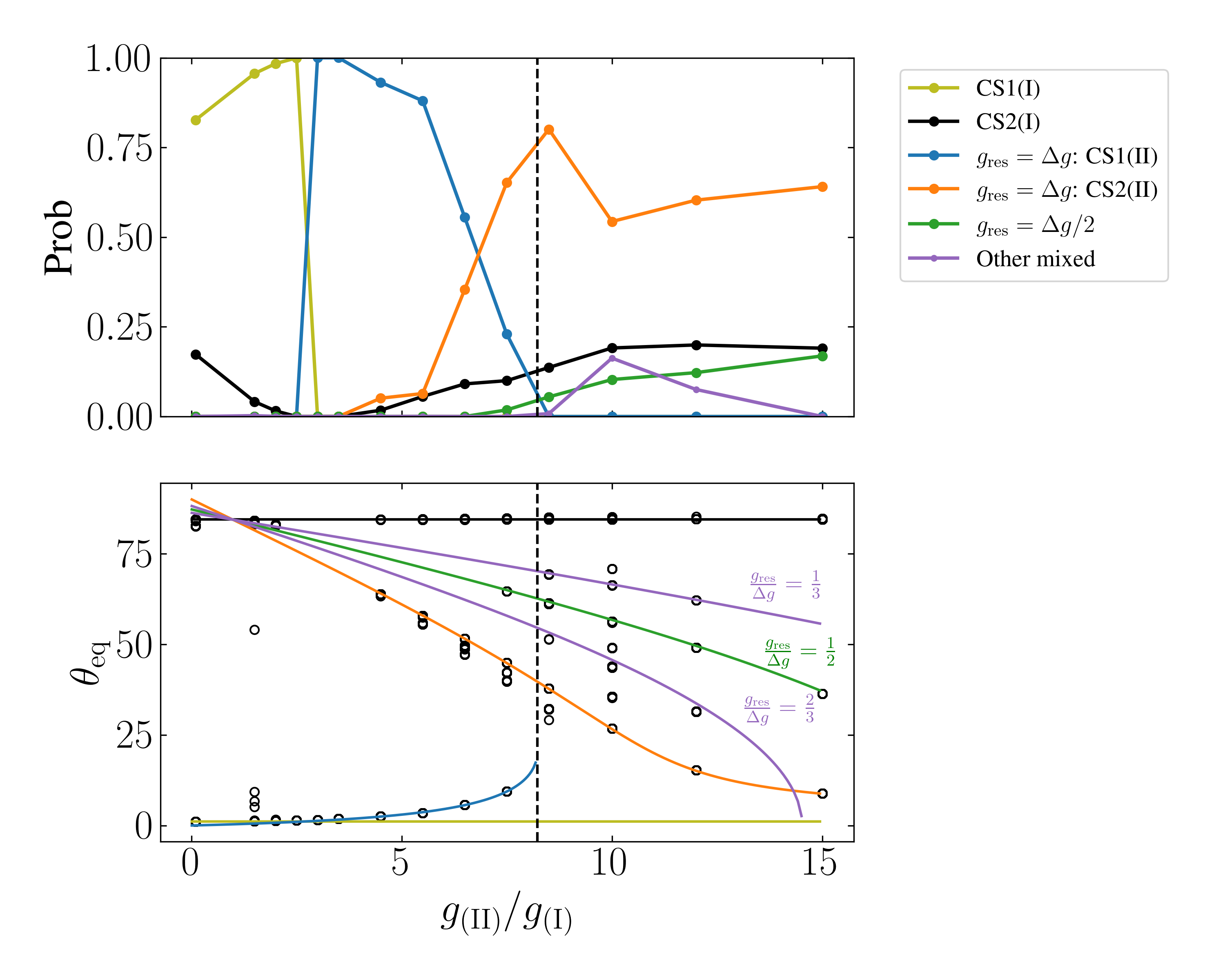}
    \caption{Same as Fig.~\ref{fig:outcomes1} but for $I_{\rm (II)} = 3^\circ$.
    Note that the agreement of the black open circles with the theoretical
    obliquities in the bottom panel is slightly worse than in
    Fig.~\ref{fig:outcomes1} but still well within the ranges of oscillation of
    the obliquities (see Fig.~\ref{fig:3outcomes010_scat} for the characteristic
    ranges). }\label{fig:outcomes3}
\end{figure*}

Having discussed how the spin evolution changes when $g_{\rm (II)}$ is varied,
we now explore the effect of different values of $I_{\rm (II)}$. In
Fig.~\ref{fig:3outcomes3}, we display the final outcomes as a function of the
initial spin orientation for the same $g_{\rm (II)}$ values as in
Fig.~\ref{fig:3outcomes1}, but for $I_{\rm (II)} = 3^\circ$. Comparing the
bottom-left panels of Figs.~\ref{fig:3outcomes1} and~\ref{fig:3outcomes3} (with
$g_{\rm (II)} = 3.5g_{\rm (I)}$), we find that the favored low-obliquity CS
changes from CS1(I) to CS1(II) when using $I_{\rm (II)} = 3^\circ$. In both
cases, CS2(I) is destabilized such that most initial conditions converge to the
low-obliquity CS, either CS1(I) or CS1(II). In the bottom-right panel, we find
that many initial conditions converge to other mixed modes than the $g_{\rm res}
= \Delta g/2$ mode. The values of $g_{\rm res}$ observed for the system are
shown in Fig.~\ref{fig:3outcomes010_scat}, where we find that many low-order
rational multiples of $g_{\rm res} / \Delta g$ are obtained. While the amplitude
of oscillation in the final $\theta_{\rm sl}$ is substantial (and larger than in
Fig.~\ref{fig:outcomes10_scat}), we find that the predictions of
Eq.~\eqref{eq:g_res_rel} are consistent up to the range of oscillation of
$\theta_{\rm sl}$. In Fig.~\ref{fig:outcomes3}, we summarize the outcomes of
spin evolution as a function of $g_{\rm (II)} / g_{\rm (I)}$ for $I_{\rm (II)} =
3^\circ$. We identify the same two qualitative trends as seen in
Fig.~\ref{fig:outcomes1}: the instability of CS2(I) when $g_{\rm (II)} \sim
g_{\rm (I)}$ and the appearance of mixed modes when $\abs{g_{\rm (II)}} \gg
\abs{g_{\rm (I)}}$.

\subsection{Summary of Various Outcomes}

In summary, the spin evolution in a 3-planet system driven by a tidal alignment
torque depends largely on the frequency of the smaller-amplitude mode, $g_{\rm
(II)}$, compared to that of the larger-amplitude mode, $g_{\rm (I)}$. In the
regime where $\abs{g_{\rm (I)}} \lesssim \alpha$, we find that:
\begin{itemize}
    \item When $\abs{g_{\rm (II)}} \ll \abs{g_{\rm (I)}}$, the low-amplitude and
        slow (II) mode does not significantly affect the spin evolution, and the
        results of \citep{su2021dynamics} are recovered. The two possible
        outcomes are the tidally stable  CS1(I) (generally low obliquity) and
        CS2(I) (generally high obliquity). Prograde initial spins converge to
        CS1(I), spins inside the mode I resonance converge to CS2(I), and
        retrograde initial spins attain one of these two tidally stable CSs
        probabilistically. For the fiducial parameters used for
        Fig.~\ref{fig:outcomes1}, approximately $20\%$ of systems are trapped in
        the high-obliquity CS2(I).

    \item When $g_{\rm (II)} \sim g_{\rm (I)}$, CS2(I) is increasingly
        difficult to attain due to the interacting mode I and mode II
        resonances, and the probability of attaining CS2(I) can be strongly
        suppressed (see Fig.~\ref{fig:outcomes1}, where the
        probability of a high-obliquity outcome goes to zero for $g_{\rm (II)} /
        g_{\rm (I)} = 3.5$).

    \item When $\abs{g_{\rm (II)}} \gtrsim \abs{g_{\rm (I)}}$, there are
        three classes of outcomes. The highest-obliquity outcome is still
        CS2(I), and is attained for initial conditions inside the mode I
        resonance (separatrix; see Fig.~\ref{fig:outcomes10_scat}). The
        lowest-obliquity outcome is generally CS2(II)\footnote{This may not be
        the case when $I_{\rm (II)} \lesssim I_{\rm (I)}$ while $g_{\rm (II)}
        \gg g_{\rm (I)}$; see Fig.~\ref{fig:3outcomes9} in the Appendix} and is
        the most favored outcome (see Fig.~\ref{fig:outcomes1}). The third
        possible outcome are mixed modes with $g_{\rm res} / \Delta g$ a
        low-order rational number (see Eq.~\ref{eq:def_phires}). These mixed
        modes only appear for $\abs{g_{\rm (II)}} \gg \abs{g_{\rm (I)}}$, and
        generally have obliquities between those of CS2(I) and CS2(II) (see
        Fig.~\ref{fig:3outcomes1}). For the fiducial parameters
        used for Fig.~\ref{fig:outcomes1}, the mixed-mode resonances increase
        the probability of obtaining a substantial ($\gtrsim 45^\circ$)
        obliquity from $\sim 20\%$ to $\sim 30\%$.
\end{itemize}

\section{Weak Tidal Friction}\label{s:weaktide}

We now briefly discuss the spin evolution of the system incorporating the full
tidal effects. In the weak friction theory of the equilibrium tide, the spin
orientation and frequency jointly evolve following \citep{alexander1973weak,
hut1981tidal, lai2012}
\begin{align}
    \p{\rd{\uv{S}}{t}}_{\rm tide} &= \frac{1}{t_{\rm s}}
                \s{\frac{2n}{\Omega_{\rm s}} - \p{\uv{S} \cdot \uv{L}}}
                    \uv{S} \times \p{\uv{L} \times \uv{S}}\label{eq:dsdt_tide},\\
    \frac{1}{\Omega_{\rm s}}\p{\rd{\Omega_{\rm s}}{t}}_{\rm tide}
        &= \frac{1}{t_{\rm s}} \s{\frac{2n}{\Omega_{\rm s}}\p{\uv{S} \cdot
            \uv{L}} - 1 - \p{\uv{S} \cdot \uv{L}}^2},\label{eq:dWsdt_tide}
\end{align}
where
\begin{align}
    \frac{1}{t_{\rm s}} &\equiv \frac{1}{4k}
        \frac{3k_2}{Q}\p{\frac{M_\star}{m}}\p{\frac{R}{a}}^3 n,
        \label{eq:ts_tide}
\end{align}
(see Eq.~\ref{eq:t_al}, but with $4k = 1$).

\begin{figure*}
    \centering
    \includegraphics[width=\columnwidth]{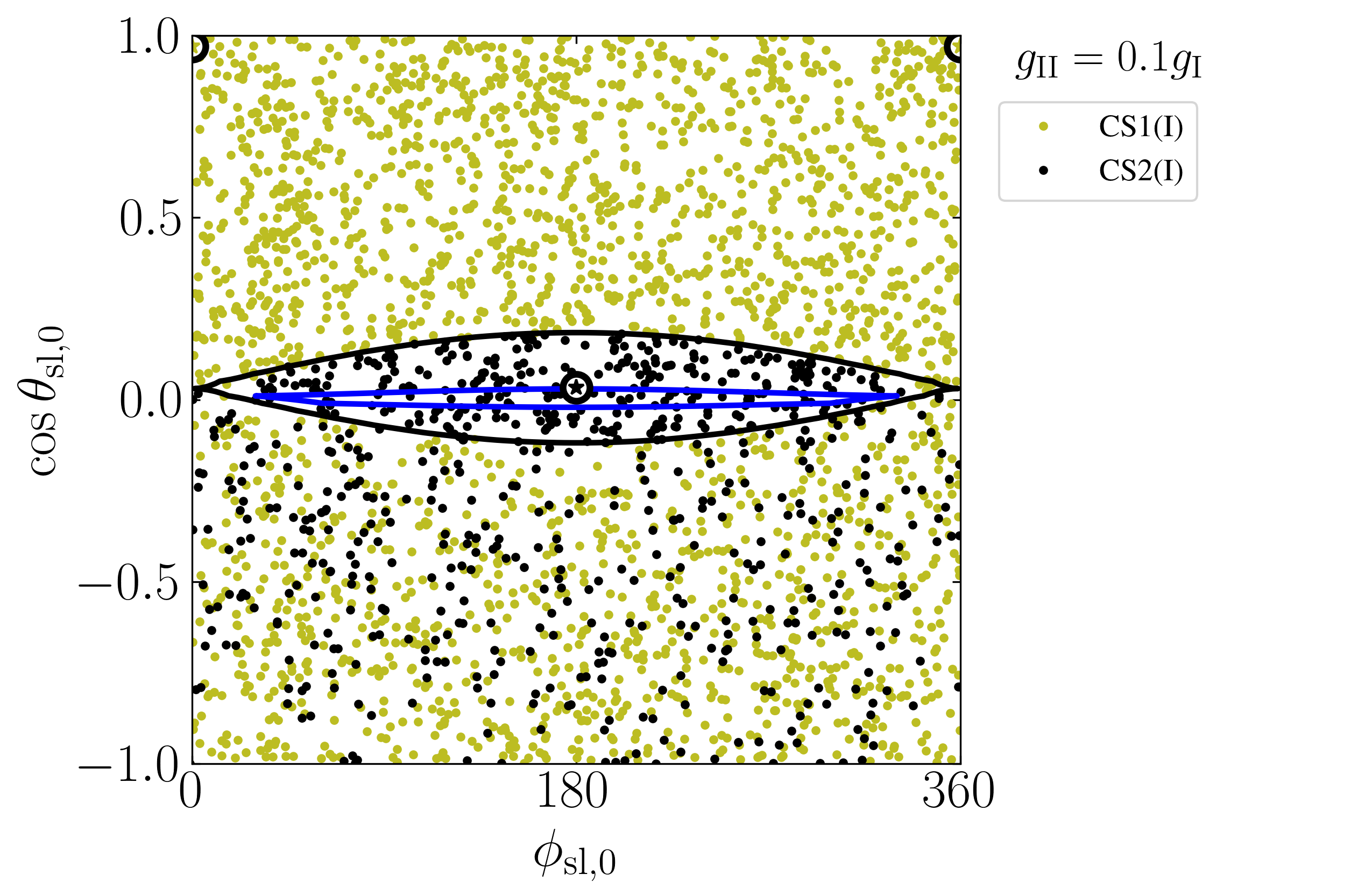}
    \includegraphics[width=\columnwidth]{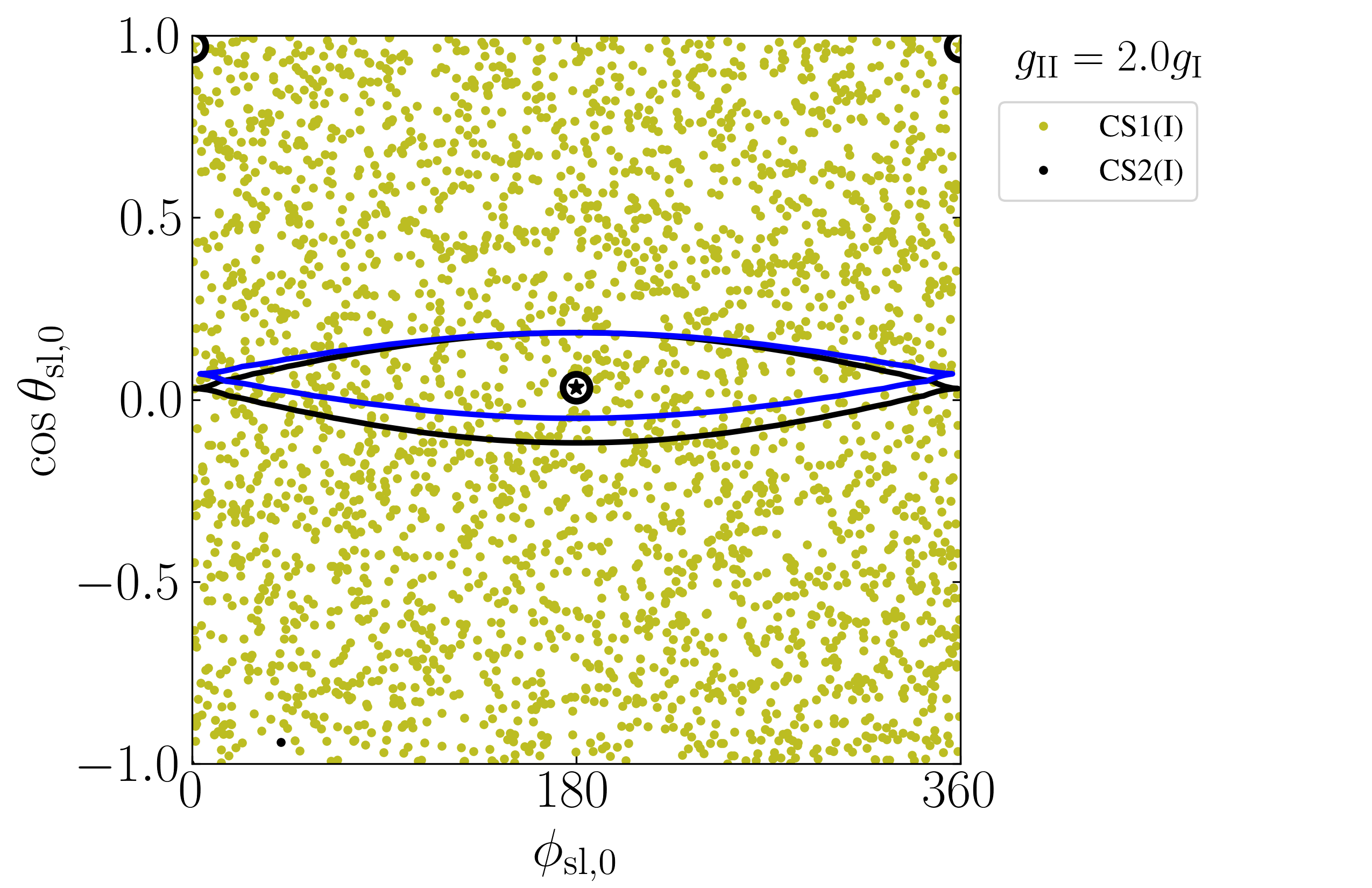}
    \includegraphics[width=\columnwidth]{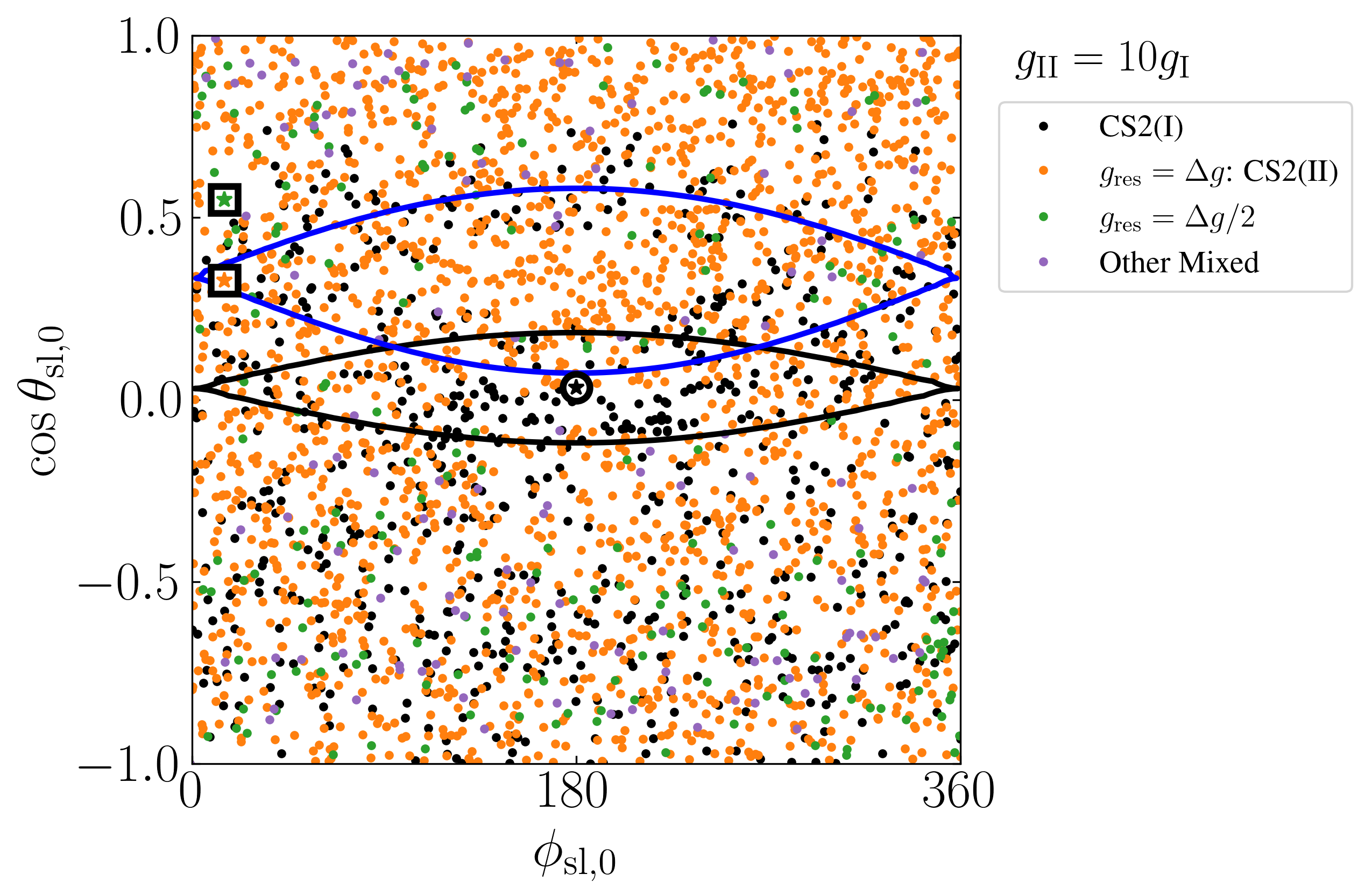}
    \caption{Similar to Figs.~\ref{fig:3outcomes3} but including full tidal
    effects on the planet's spin, with $\alpha_{\rm sync} = 10\abs{g_{\rm (I)}}$
    (Eq.~\ref{eq:def_wslsync}), $I_{\rm (II)} 3^\circ$, and initial spin
    $\Omega_{\rm s, 0} = 3n$. In the three panels, $g_{\rm (II)} = \z{0.1, 2,
    10} \times g_{\rm (I)}$ respectively. Note that the plotted separatrices
    (blue and black lines) use the initial value of $\alpha$. }\label{fig:wt}
\end{figure*}

\begin{figure*}
    \centering
    \includegraphics[width=0.8\textwidth]{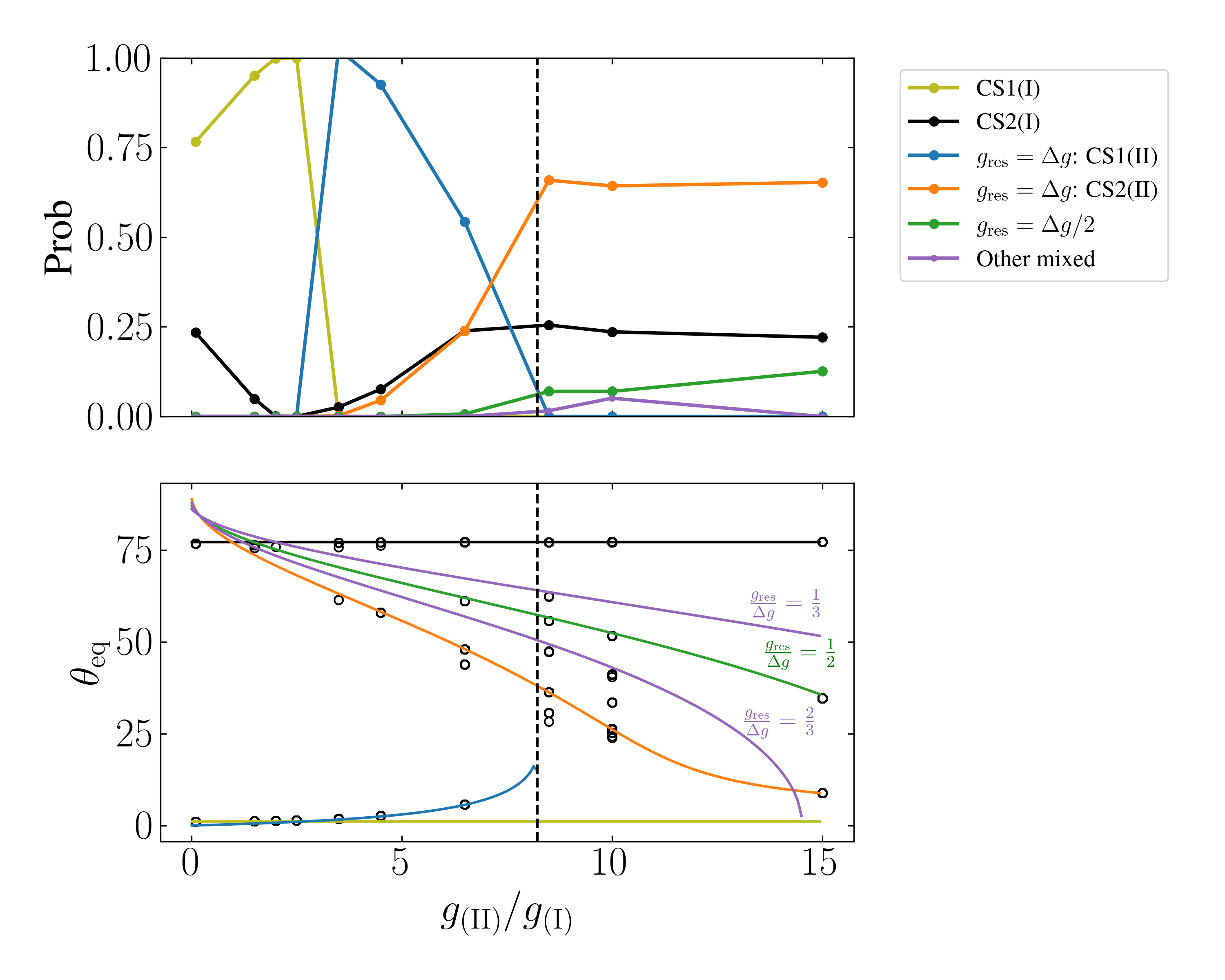}
    \caption{Similar to Fig.~\ref{fig:outcomes3} but with weak tidal
    friction.}\label{fig:wt_outcomes}
\end{figure*}

Since $\alpha \propto \Omega_{\rm s}$ evolves in time, we describe the
spin-orbit coupling by the parameter
\begin{equation}
    \alpha_{\rm sync} \equiv \s{\alpha}_{\Omega_{\rm s} = n}.
        \label{eq:def_wslsync}
\end{equation}
To facilitate comparison with the previous results, we use $\alpha_{\rm sync} =
10\abs{g_{\rm (I)}}$ and $\abs{g_{\rm (I)}}t_{\rm s} = 300$. Note that for the
physical parameters used in Eqs.~\eqref{eq:wsl} and~\eqref{eq:ts_tide}, $g_{\rm
(I)}t_{\rm s} \sim 10^4$; we choose a faster tidal timescale to accelerate our
numerical integrations. The initial spin is fixed $\Omega_{\rm s, 0} = 3n$.
We then integrate Eqs.~\eqref{eq:dsdt1},~\eqref{eq:l_t_sol},
and~(\ref{eq:dsdt_tide}--\ref{eq:dWsdt_tide}) starting from various initial spin
orientations and determine the final outcomes. Figure~\ref{fig:wt} shows the
results for a few select values of $g_{\rm (II)}$ and for $I_{\rm (II)} =
3^\circ$. Similar behaviors to Figs.~\ref{fig:3outcomes1}
and~\ref{fig:3outcomes3} are observed. The probabilities and
obliquities of the various equilibria are shown in
Fig.~\ref{fig:wt_outcomes}. Note that each equilibrium obliquity has a
corresponding equilibrium rotation rate, given by
\begin{equation}
    \frac{\Omega_{\rm eq}}{n} = \frac{2 \cos \theta_{\rm eq}}{
        1 + \cos^2\theta_{\rm eq}}.
\end{equation}
The probabilities shown in Fig.~\ref{fig:wt_outcomes} exhibit qualitative trends
that are quite similar to those seen for the evolution driven by the tidal
alignment torque alone: when $\abs{g_{\rm (II)}} \ll \abs{g_{\rm (I)}}$, the
results of \citep{su2021dynamics} are recovered; when $g_{\rm (II)} \sim g_{\rm
(I)}$, the probability of attaining CS2(I) is significantly suppressed; and when
$\abs{g_{\rm (II)}} \gg \abs{g_{\rm (I)}}$, mixed modes appear.

\section{Summary and Discussion}\label{s:disc}

In this work, we have shown that the planetary spins in compact systems of
multiple super-Earths (SEs), possibly with an outer cold Jupiter companion, can
be trapped into a number of spin-orbit resonances when evolving under tidal
dissipation, either via a tidal alignment torque (Section~\ref{s:align}) or via
weak tidal friction (Section~\ref{s:weaktide}). In addition to the
well-understood tidally-stable Cassini States associated with each of the
orbital precession modes, we have also discovered a new class of ``mixed mode''
spin-orbit resonances that yield substantial obliquities. These additional
resonances constitute a significant fraction of the final states of tidal
evolution if the planet's initial spin orientation is broadly distributed, a
likely outcome for planets that have experienced an early phase of collisions or
giant impacts. For instance, for the fiducial system parameters
shown in Fig.~\ref{fig:outcomes1}, these mixed-mode equilibria increase the
probability that a planet retains a substantial ($\gtrsim 45^\circ$) obliquity
from $20\%$ to $30\%$. A large equilibrium obliquity has a significant
influence on the planet's insolation and climate. For planetary systems
surrounding cooler stars (M dwarfs), the SEs (or Earth-mass planets) studied in
this work lie in the habitable zone \citep[e.g.][]{dressing2017,
gillon2017seven}, and the nontrivial obliquity can impact the habitability of
such planets.

In a broader sense, the mixed-mode equilibria discovered in our study represent
a novel astrophysical example of subharmonic responses in parametrically
driven nonlinear oscillators. In equilibrium, the planetary
obliquity oscillates with a period that is an integer multiple of the driving
period $2\pi / \abs{\Delta g}$ (see Appendix~\ref{app:mixed_mode} for further
discussion). Such subharmonic responses are often seen in nonlinear oscillators
\citep[e.g.\ in the classic van der Pol and Duffing equations][]{
levenson1949harmonic, flaherty1978frequency, hayashi2014nonlinear}.

Throughout this paper, we have adopted fiducial parameters where $\abs{g_{\rm
(I)}} \lesssim \alpha$, which is generally expected for the SE systems being
studied. If instead $\abs{g_{\rm (I)}} \gtrsim \alpha$, then there is no
resonance for the dominant (larger-amplitude) mode I. There are then a few
possible cases: if $\abs{g_{\rm (II)}} \lesssim \alpha$, mode II is both slower
than mode I and has a smaller amplitude, so it will not affect the mode I
dynamics significantly. On the other hand, if $\abs{g_{\rm (II)}} \gtrsim
\alpha$, then mode II also has no resonance, and both CS2(I) and CS2(II) have
low obliquities, implying that the system will always settle into a
low-obliquity state.

\section*{Acknowledgements.}

This work has been supported in part by the NSF grant AST-17152 and NASA grant
80NSSC19K0444. YS is supported by the NASA FINESST grant
19-ASTRO19-0041. 

\section*{Data Availability}

The data referenced in this article will be shared upon reasonable request to
the corresponding author.

\clearpage

\bibliography{Su_multi_cs}
\bibliographystyle{mnras}

\appendix

\clearpage

\counterwithin{figure}{section}

\section{Materials and Methods}

\subsection{Inclination Modes of 3-Planet Systems}\label{app:l_evol}

In the linear regime, the evolution of the orbital inclinations in a
multi-planet system is described by the Laplace-Lagrange
theory \citep{murray1999solar, pu2018}. In this section, we consider 3-planet
systems. We denote the magnitude of the angular momentum of each planet by $L_j$
and the inclination relative to the total angular momentum axis $\uv{J}$ by
$I_j$, and we define the complex inclination $\mathcal{I}_j =
I_j\exp\s{i\Omega_{i}}$. The evolution equations for $\mathcal{I}_1$,
$\mathcal{I}_2$, and $\mathcal{I}_3$ are
\begin{equation}
    \rd{}{t}\begin{pmatrix}
        \mathcal{I}_1\\ \mathcal{I}_2\\\mathcal{I}_3
    \end{pmatrix}
        = i \begin{pmatrix}
            -\omega_{12} - \omega_{13} & \omega_{12} & \omega_{13}\\
            \omega_{21} & -\omega_{21} - \omega_{23} & \omega_{23}\\
            \omega_{31} & \omega_{32} & -\omega_{31} - \omega_{32},
        \end{pmatrix}
    \begin{pmatrix}
        \mathcal{I}_1\\ \mathcal{I}_2\\\mathcal{I}_3
    \end{pmatrix},\label{eq:app_inc_eom}
\end{equation}
where $\omega_{jk}$ is the precession rate of the $j$-th planet induced by the
$k$-th planet, and is given by
\begin{equation}
    \omega_{jk} = \frac{m_k}{4M_\star}
        \frac{a_j a_<}{a_>^2}
        n_j b_{3/2}^{(1)}(\alpha),
\end{equation}
where $a_< = \min\p{a_j, a_k}$, $a_> = \max\p{a_j, a_k}$, $n_j = (GM_\star /
a_j^3)^{1/2}$, $\alpha = a_< / a_>$, and
\begin{equation}
    b_{3/2}^{(1)}(\alpha) = 3\alpha\p{
        1 + \frac{15}{8}\alpha^2 + \frac{175}{64}\alpha^4 + \dots}
\end{equation}
is the Laplace coefficient. Eq.~\eqref{eq:app_inc_eom} can be solved in general,
giving two non-trivial eigenmodes. In the limit $L_1 \ll L_2, L_3$, the two
eigenmodes have simple solutions and interpretations:
\begin{itemize}
    \item Mode I has frequency
        \begin{equation}
            g_{\rm (I)} = -\p{\omega_{12} + \omega_{13}}.\label{eq:Ig_g1}
        \end{equation}
        It corresponds to free precession of $\uv{L}_1$ around the total angular
        momentum $\bm{J} = \bm{L}_2 + \bm{L}_3$. The amplitude of
        oscillation of $\uv{L}_1$, $I_{\rm (I)}$, is simply the inclination of
        $\uv{L}_1$ with respect to $\uv{J}$.

    \item Mode II has frequency
        \begin{equation}
            g_{\rm (II)} = -\p{\omega_{23} + \omega_{32}} =
            -\frac{J}{L_3}\omega_{23},\label{eq:Ig_g2}
        \end{equation}
        which is simply the precession frequency of $\uv{L}_2$ (or
        $\uv{L}_3$) about $\uv{J}$. The forced oscillation of $\uv{L}_1$ has an
        amplitude
        \begin{equation}
            I_{\rm (II)} =
                \frac{\omega_{12}L_3 - \omega_{13}L_2}
                {\p{g_{\rm (II)} - g_{\rm (I)}}J} I_{23},
                \label{eq:Ig_mode2}
        \end{equation}
        where $I_{23}$ is the mutual inclination between the two outer planets
        and is constant.
\end{itemize}

\begin{figure}
    \centering
    \includegraphics[width=\columnwidth]{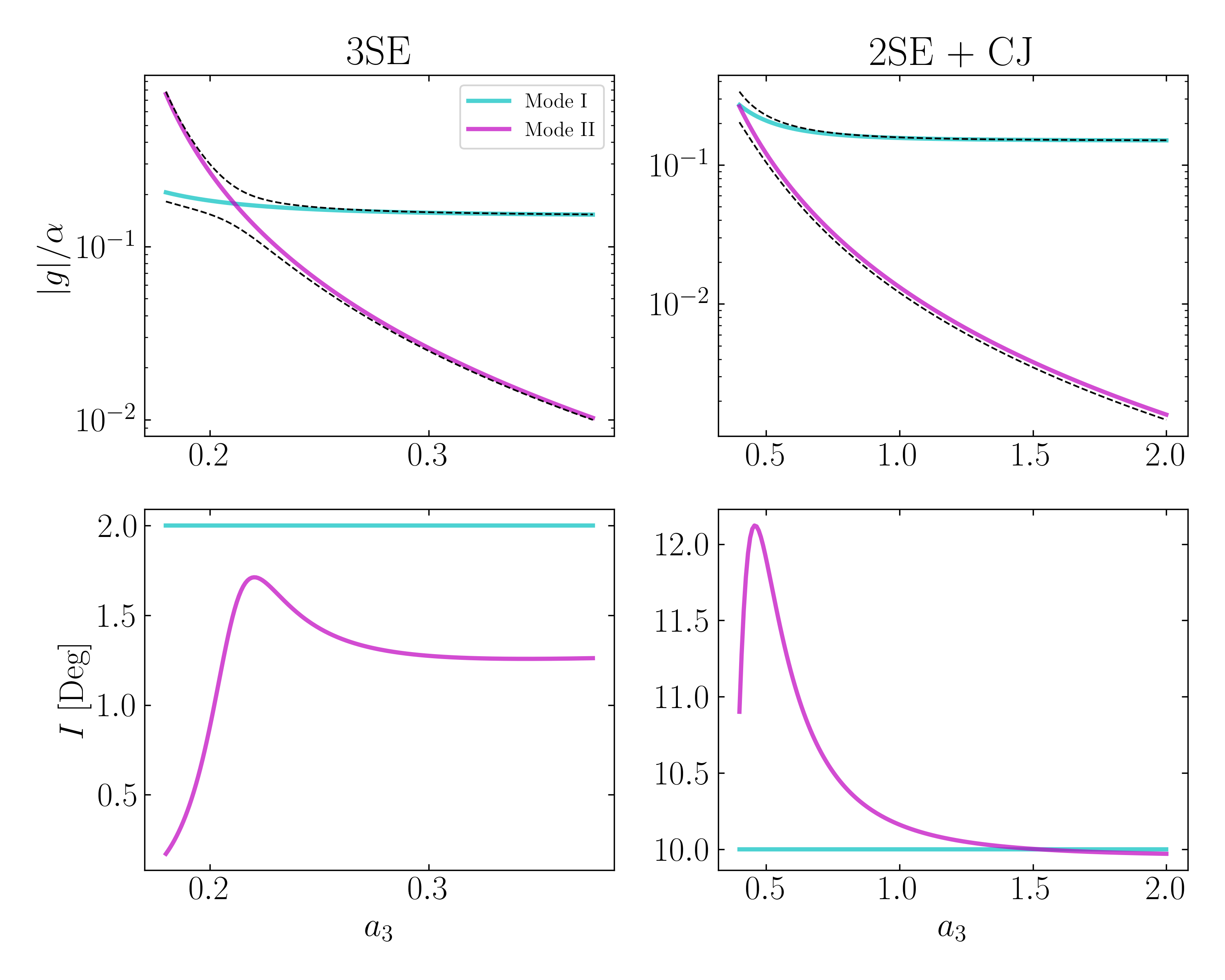}
    \caption{Inclination mode frequencies and amplitudes for the 3SE (left) and
    2SE + CJ (right) systems. In both systems, the inner planets' parameters are
    $m_1 = M_{\oplus}$, $m_2 = 3M_{\oplus}$, $a_1 = 0.1\;\mathrm{AU}$, $a_2 =
    0.15\;\mathrm{AU}$. In the 3SE case, $m_3 = 3M_{\oplus}$ and $I_1 = I_{23} =
    2^\circ$, while in the 2SE + CJ case, $m_3 = 0.5M_{\rm J}$ and $I_1 = I_{23}
    = 10^\circ$, In the top panels, the black dashed lines show the mode
    frequencies from the exact solution of Eq.~\eqref{eq:app_inc_eom}, while the
    solid, colored lines are given by
    Eqs.~(\ref{eq:Ig_g1},~\ref{eq:Ig_g2}).}\label{fig:Ig_sweep}
\end{figure}

\begin{figure}
    \centering
    \includegraphics[width=\columnwidth]{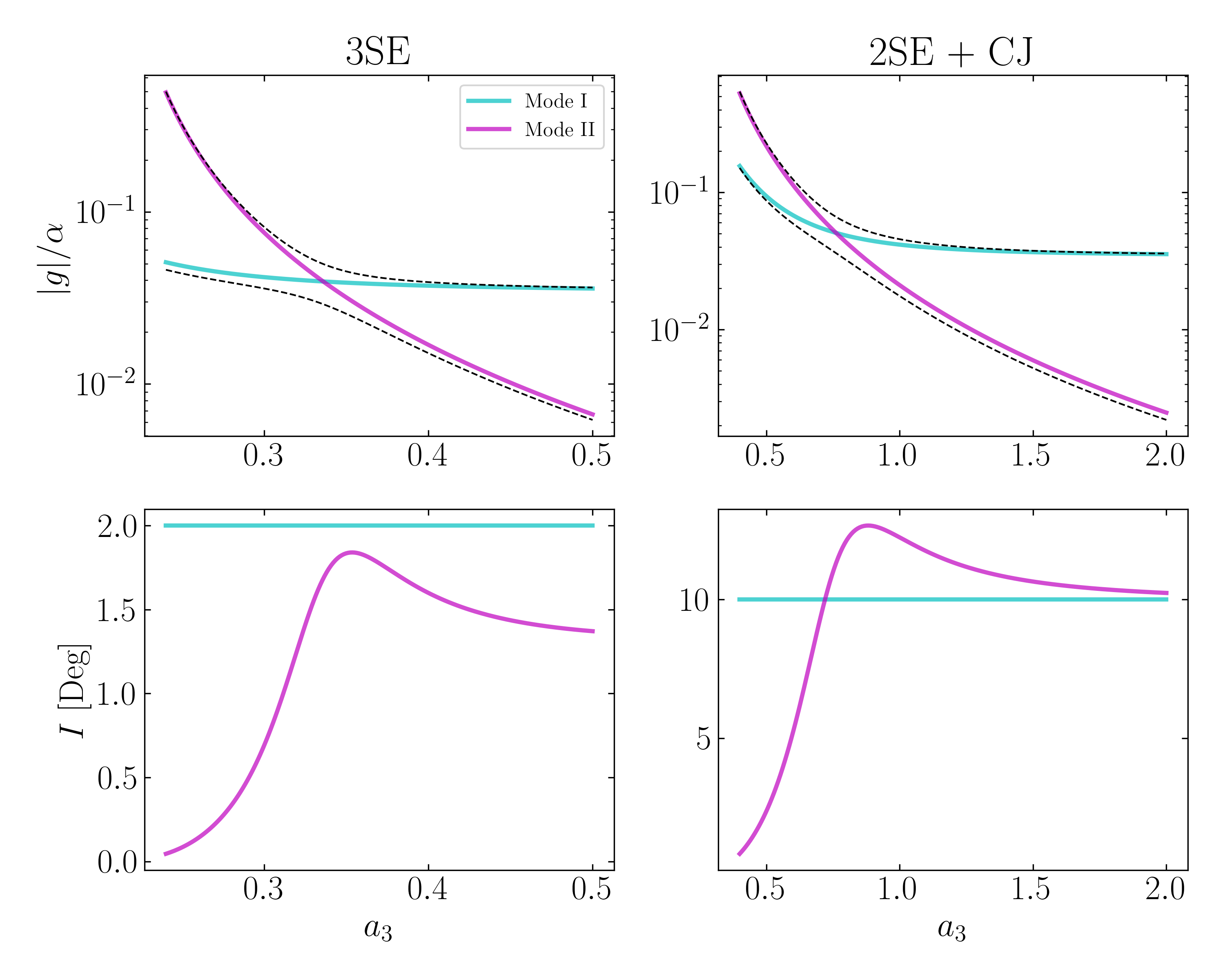}
    \caption{Same as Fig.~\ref{fig:Ig_sweep} but for $a_2 =
    0.2\;\mathrm{AU}$.}\label{fig:Ig_sweep2}
\end{figure}

\begin{figure}
    \centering
    \includegraphics[width=\columnwidth]{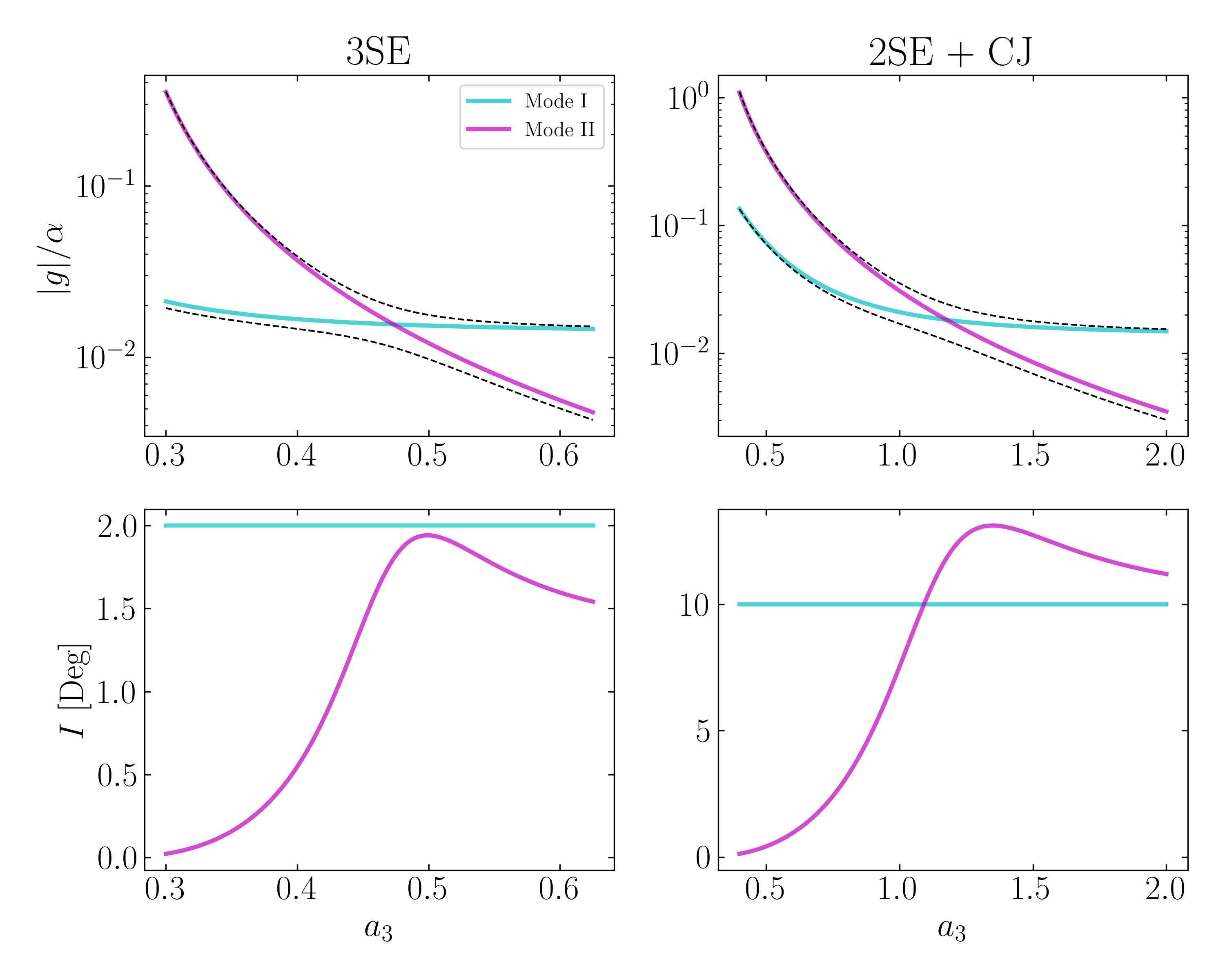}
    \caption{Same as Fig.~\ref{fig:Ig_sweep} but for $a_2 =
    0.25\;\mathrm{AU}$.}\label{fig:Ig_sweep25}
\end{figure}

We consider two archetypal 3-planet configurations, systems with three super
Earths (SEs) and systems with two inner SEs and an exterior cold Jupiter (CJ).
In both cases, we take the inner two planets to have $m_1 = M_{\oplus}$, $m_2 =
3M_{\oplus}$, $a_1 = 0.1\;\mathrm{AU}$, and we consider three values of $a_2 =
\z{0.15, 0.2, 0.25}\;\mathrm{AU}$. For the 3SE case, we take $m_3 = 3M_{\oplus}$
and the characteristic inclinations $I_1 \simeq I_{23} \simeq 2^\circ$
[corresponding to three nearly-coplanar SEs; see \citealp{fabrycky_mutual_incs,
dai2018larger}]. For the 2SE + CJ case, we take $m_3 = 0.5M_{\rm J}$ and the
characteristic inclinations $I_1 \simeq I_{23} \simeq 10^\circ$ [corresponding
to a mildly inclined CJ\@; see \citealp{masuda2020mutual}]. In both cases, we
compute the mode precession frequencies $g_{\rm (I, II)}$ and characteristic
mode amplitudes $I_{\rm (I, II)}$ for a range of $a_3$.
Figures~\ref{fig:Ig_sweep}--\ref{fig:Ig_sweep25} show examples of our results.

\subsection{Additional Comments and Results on Mixed Mode
Equilibria}\label{app:mixed_mode}

\begin{figure*}
    \centering
    \includegraphics[width=1.3\columnwidth]{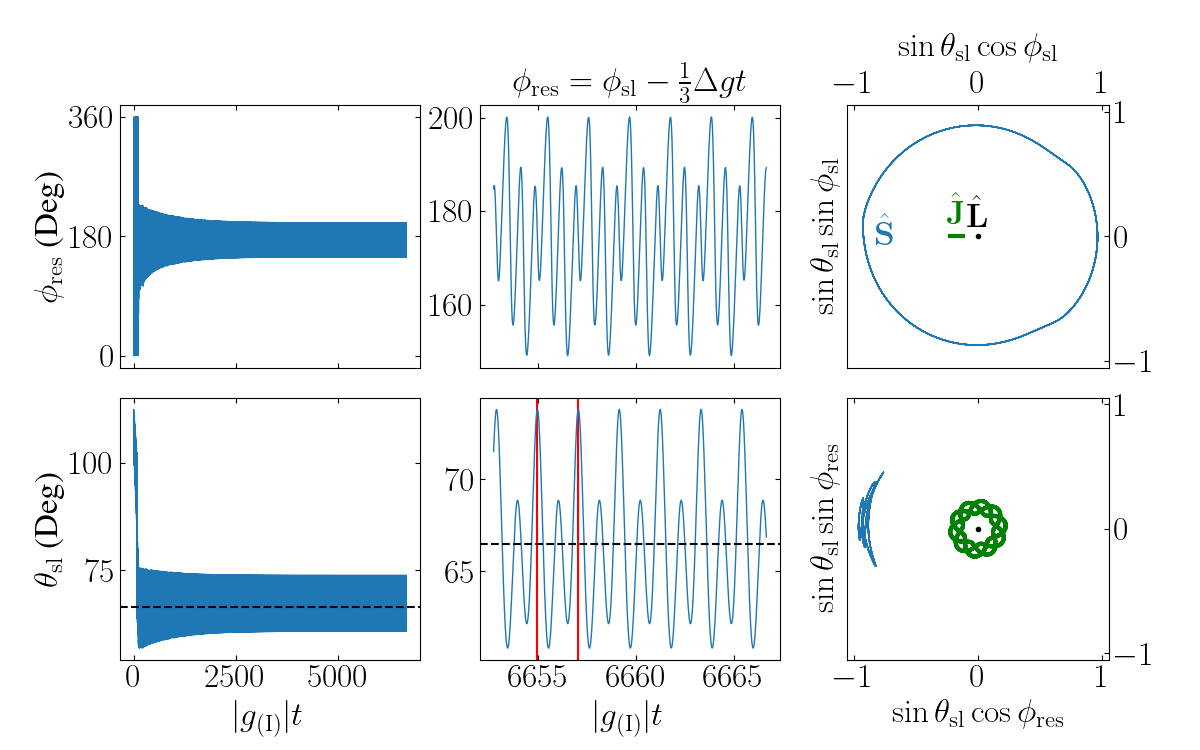}
    \includegraphics[width=1.3\columnwidth]{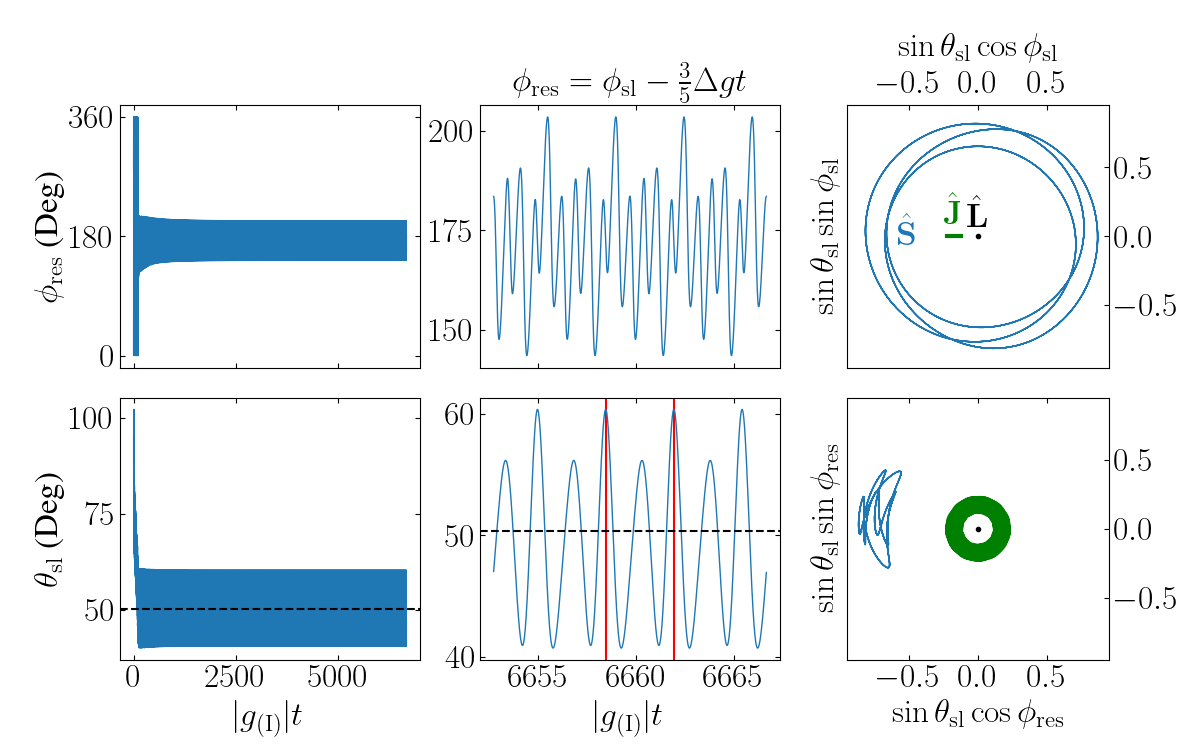}
    \includegraphics[width=1.3\columnwidth]{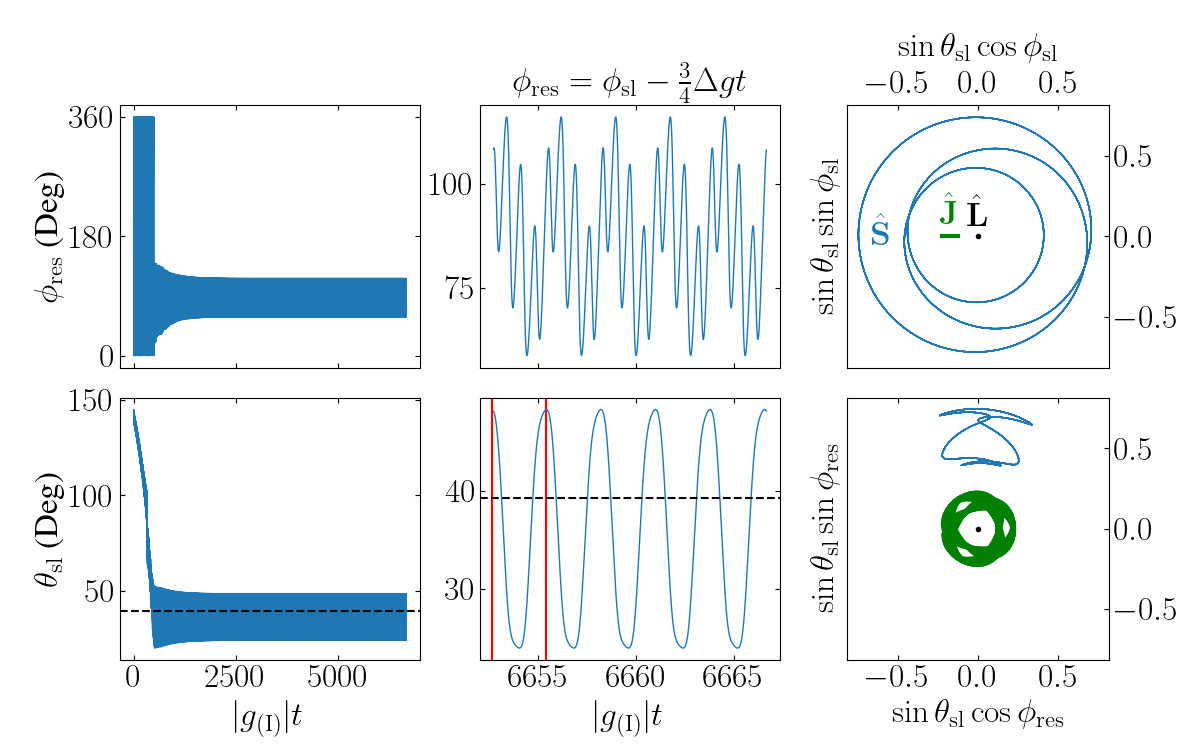}
    \caption{Same as Fig.~\ref{fig:example1_10} but for $g_{\rm (II)} = 10g_{\rm
    (I)}$ and $I_{\rm (II)} = 3^\circ$. The three examples correspond to capture
    into mixed-mode equilibria with resonant angles corresponding to $g_{\rm
    res} = \Delta g / 3$, $g_{\rm res} = 3\Delta g / 5$, and $g_{\rm res} =
    3\Delta g / 4$ respectively. The three pairs of vertical red lines are
    separated by $6\pi / \abs{\Delta g}$, $10\pi / \abs{\Delta g}$, and $8\pi /
    \Delta g$ in the three examples respectively.}\label{fig:example3_10}
\end{figure*}

In the main text, we provided an example of the spin evolution into a mixed-mode
equilibrium in Fig.~\ref{fig:example1_10} for the parameters $I_{\rm (I)} =
10^\circ$, $I_{\rm (II)} = 1^\circ$, $\alpha = 10\abs{g_{\rm (I)}}$, and $g_{\rm
(II)} = 10g_{\rm (I)}$. In Fig.~\ref{fig:example3_10}, we provide several
further examples of the evolution into other mixed-mode equilibria with
different values of $g_{\rm res}$ when $I_{\rm (II)} = 3^\circ$ is used. We find
that their average equilibrium obliquities $\theta_{\rm eq}$ are still
well-described by Eq.~\eqref{eq:g_res_rel}. Furthermore, we find that, if
$g_{\rm res} / \Delta g = p / q$ for integers $p$ and $q$, then the steady-state
oscillations of $\theta_{\rm sl}$ are periodic with period $2\pi q / \abs{\Delta
g}$ (see bottom example in Fig.~\ref{fig:example1_10} for the case of $g_{\rm
res} = \Delta g / 2$).

\begin{figure}
    \centering
    \includegraphics[width=\columnwidth]{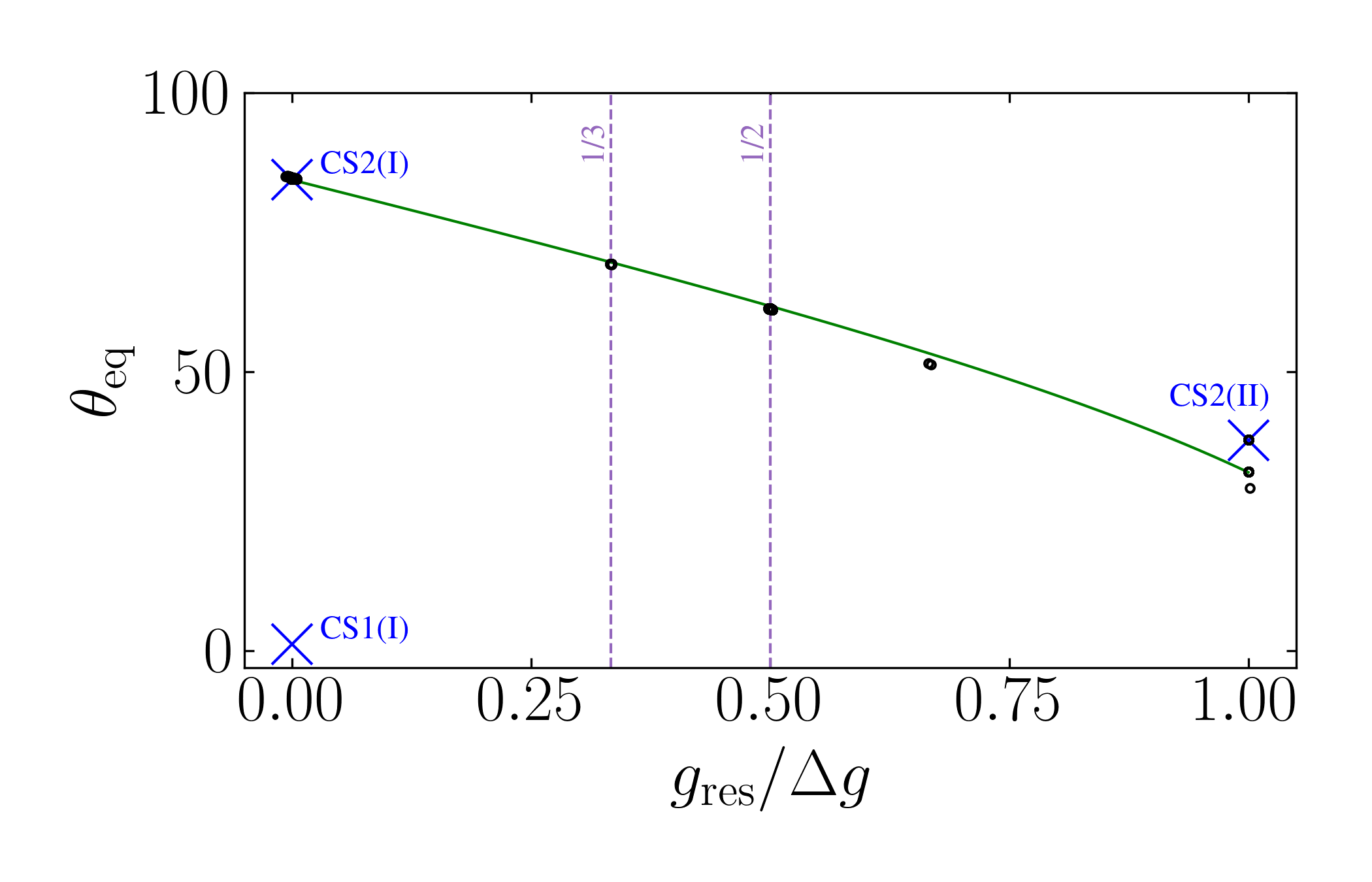}
    \caption{Same as Figs.~\ref{fig:outcomes10_scat}
    and~\ref{fig:3outcomes010_scat} but for $I_{\rm (II)} = 3^\circ$ and $g_{\rm
    (II)} = 8.5g_{\rm (I)}$, where ranges of oscillation in $\theta_{\rm sl}$
    have been suppressed for clarity.}\label{fig:3outcomes85_scat}
\end{figure}

Figure~\ref{fig:3outcomes85_scat} shows the equilibrium obliquity $\theta_{\rm
eq}$ as a function of $g_{\rm res}$ for a system with $I_{\rm (II)} = 3^\circ$
and $g_{\rm (II)} = 8.5g_{\rm (I)}$. We can see for $g_{\rm res} = \Delta g$
that there are three distinct equilibrium values of $\theta_{\rm sl}$. The
largest-obliquity equilibrium is CS2(II), and the equilibrium with an
intermediate obliquity is a mixed-mode equilibrium with $g_{\rm res} = \Delta
g$, as it directly intersects the green line (Eq.~\ref{eq:g_res_rel}). The
existence and stability of this equilibrium is responsible for the extra dot a
few degrees below the CS2(II) curve in the bottom panels of
Figs.~\ref{fig:outcomes1},~\ref{fig:outcomes3}, and~\ref{fig:wt_outcomes} (e.g.\
most visible for $g_{\rm (II)} / g_{\rm (I)} = 6.5$, $7.5$, and $8.5$ in
Fig.~\ref{fig:outcomes3}). The origin of the lowest-obliquity equilibrium at
$g_{\rm res} = \Delta g$ in Fig.~\ref{fig:3outcomes85_scat} is distinct, though
it is within the range of oscillation of $\theta_{\rm sl}$ of the mixed-mode
steady state.

\begin{figure*}
    \centering
    \includegraphics[width=1.3\columnwidth]{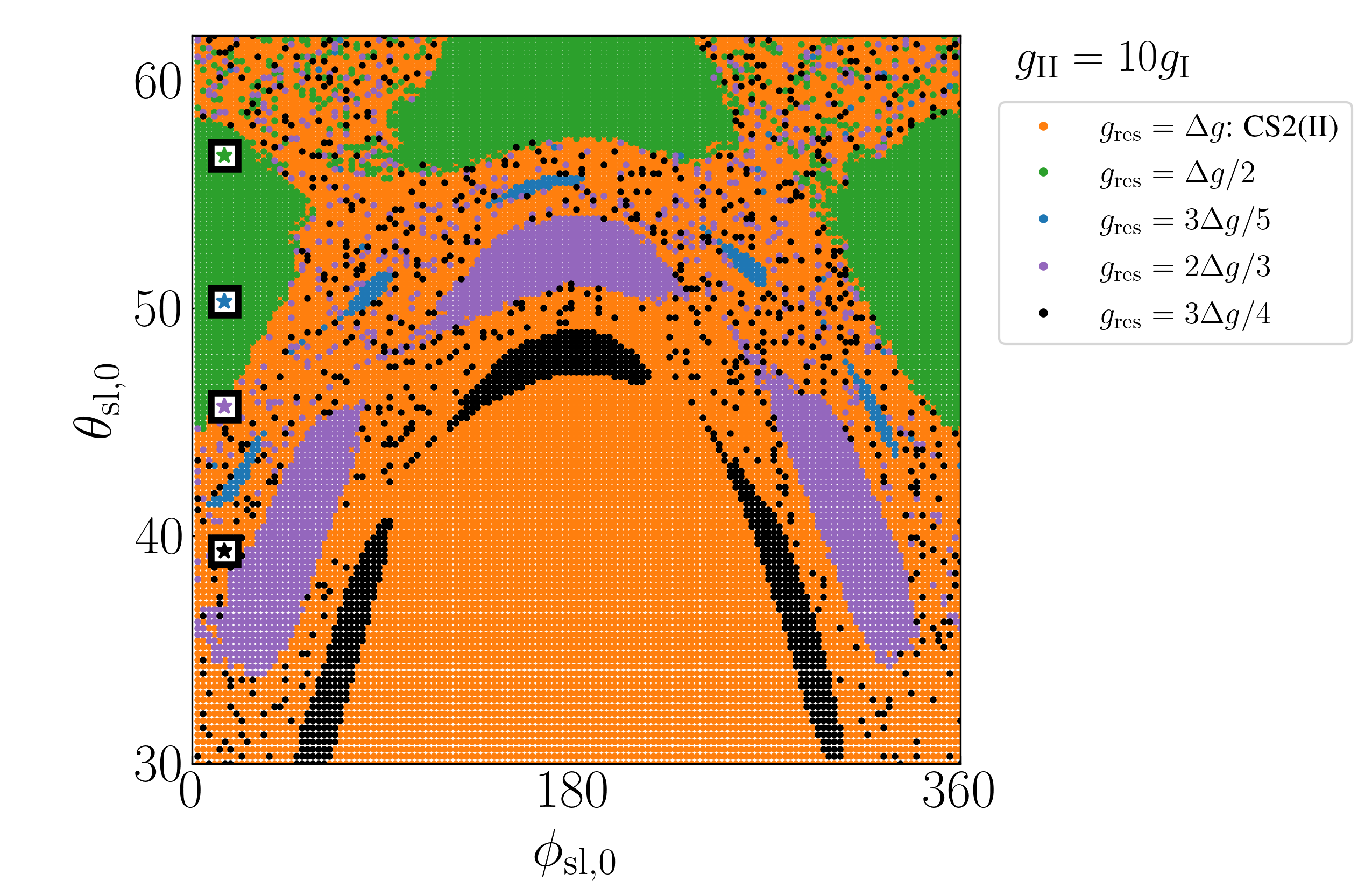}
    \caption{Same as Fig.~\ref{fig:outcomes_grid} but for $I_{\rm (II)} =
    3^\circ$. The range of the vertical axis is chosen to include the ranges of
    oscillation of the $1/2$, $3/5$, $2/3$, and $3/4$ mixed mode resonances (see
    Fig.~\ref{fig:3outcomes010_scat}). The decreasing density of points as
    $\theta_{\rm sl, 0}$ decreases is because the grid of initial conditions is
    uniform in $\cos \theta_{\rm sl, 0}$ rather than $\theta_{\rm sl, 0}$
    itself. }\label{fig:3outcomes_grid}
\end{figure*}

In Fig.~\ref{fig:outcomes_grid}, we presented the numerical stability analysis
of initial conditions in the neighborhood of the mixed-mode equilibrium for the
$I_{\rm (II)} = 1^\circ$ case. When $I_{\rm (II)}$ is increased to $3^\circ$,
the amplitudes of oscillations of the mixed-mode equilibria begin to overlap
(see Fig.~\ref{fig:3outcomes010_scat}), and so we might expect that the basins
of attraction for the resonances overlap in $\theta_{\rm sl, 0}$ space.
Figure~\ref{fig:3outcomes_grid} shows that this is indeed the case, and that the
basins of attraction of the mixed modes are very distorted (likely due to
interactions among the resonances) compared to those seen in the $I_{\rm (II)} =
1^\circ$ case.

\begin{figure*}
    \centering
    \includegraphics[width=0.8\textwidth]{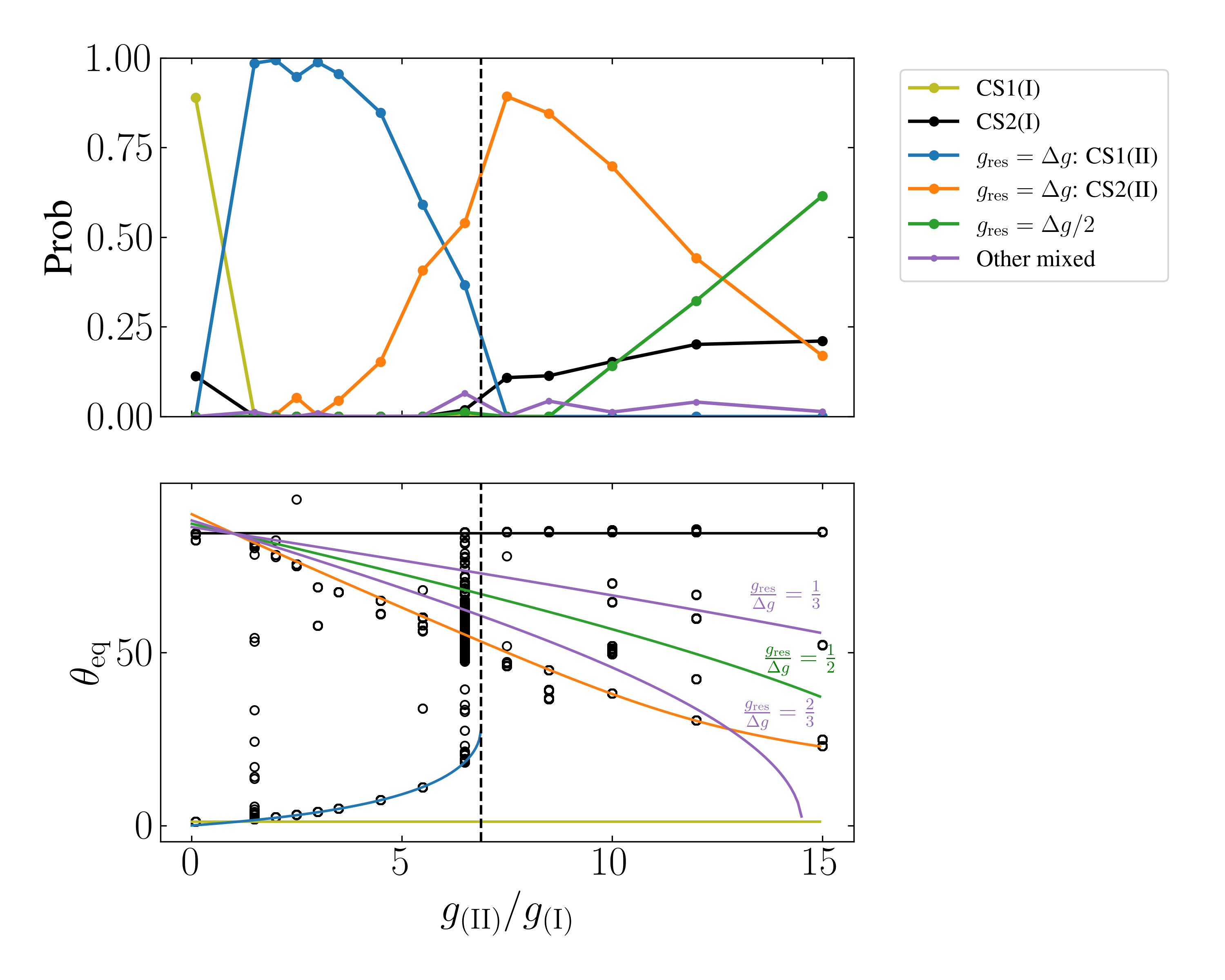}
    \caption{Same as Fig.~\ref{fig:outcomes1} but for $I_{\rm (II)} =
    9^\circ$.}\label{fig:3outcomes9}
\end{figure*}

Finally, in the main text, we have focused on the regime where $I_{\rm (II)} \ll
I_{\rm (I)}$. In Fig.~\ref{fig:3outcomes9}, we show the effect of choosing
$I_{\rm (II)} = 9^\circ$ (with the same that $I_{\rm (I)} = 10^\circ$). It can
be seen that the $g_{\rm res} = \Delta g / 2$ mixed mode is the most common
outcome when $\abs{g_{\rm (II)}} \gg \abs{g_{\rm (I)}}$, as it is the preferred
low-obliquity equilibrium ($\theta_{\rm eq} \lesssim 20^\circ$) for $g_{\rm
(II)} = 15g_{\rm (I)}$. The degraded agreement of the $\theta_{\rm eq}$ values
with Eq.~\eqref{eq:g_res_rel} is because our theoretical results assume $I_{\rm
(II)} \ll I_{\rm (I)}$. A broad range of final obliquities is observed when
$g_{\rm (II)} = 6.5g_{\rm (I)}$, very close to the critical value of $g_{\rm
(II)}$ (denoted by the vertical dashed line) where the number of mode II CSs
changes from 4 to 2 \citep{su2021dynamics}. This is likely due to the unusual
phase space structure near this bifurcation.

\label{lastpage} 
\end{document}